\documentclass[twocolumn]{aastex62}
\usepackage{amsmath}
\usepackage{amssymb}
\usepackage{mathbbol}
\usepackage{dsfont}
\usepackage{amsfonts}
\usepackage{color}
\usepackage{amsopn}
\usepackage{graphicx,longtable,times,threeparttable,multirow}
\usepackage{longtable,times,threeparttable,multirow}
\usepackage{float}

\DeclareGraphicsExtensions{.eps,.eps.gz,.epsi, .jpg, .png}

\shorttitle{Ba-rich stars}
\shortauthors{Xiang et al.}

\begin{document}

\title{Chemically peculiar A and F stars with enhanced s-process and iron-peak elements: \\ stellar radiative acceleration at work}

\author{Mao-Sheng Xiang}
\affil{Max-Planck Institute for Astronomy, K\"onigstuhl 17, D-69117 Heidelberg, Germany}
\email{{\rm email: } {\em mxiang@mpia.de}}
\author{Hans-Walter Rix}
\affil{Max-Planck Institute for Astronomy, K\"onigstuhl 17, D-69117 Heidelberg, Germany}
\author{Yuan-Sen Ting}\thanks{Hubble fellow}
\affil{Institute for Advanced Study, Princeton, NJ 08540, USA}
\affil{Department of Astrophysical Sciences, Princeton University, Princeton, NJ 08544, USA}
\affil{Observatories of the Carnegie Institution of Washington, 813 Santa
Barbara Street, Pasadena, CA 91101, USA}
\affil{Research School of Astronomy \& Astrophysics, Australian National University, Canberra, ACT 2611, Australia}
\author{Hans-Günter Ludwig}
\affil{ZAH - Landessternwarte, Universität Heidelberg, Königstuhl 12, 69117
Heidelberg, Germany}
\author{Johanna Coronado}
\affil{Max-Planck Institute for Astronomy, K\"onigstuhl 17, D-69117 Heidelberg, Germany}
\author{Meng Zhang}
\affil{Department of Astronomy, School of Physics, Peking University, Beijing 100871, P. R. China}
\affil{Kavli institute of Astronomy and Astrophysics, Peking University, Beijing 100871, P. R. China}
\author{Hua-Wei Zhang}
\affil{Department of Astronomy, School of Physics, Peking University, Beijing 100871, P. R. China}
\affil{Kavli institute of Astronomy and Astrophysics, Peking University, Beijing 100871, P. R. China}
\author{Sven Buder}
\affil{Max-Planck Institute for Astronomy, K\"onigstuhl 17, D-69117 Heidelberg, Germany}
\affil{Research School of Astronomy \& Astrophysics, Australian National University, ACT 2611, Australia}
\affil{Center of Excellence for Astrophysics in Three Dimensions (ASTRO-3D), Australia}
\author{Piero Dal Tio}
\affil{Dipartimento di Fisica e Astronomia Galileo Galilei, Università di Padova, Vicolo dell’Osservatorio 3, 35122 Padova, Italy}
\affil{Osservatorio Astronomico di Padova – INAF, Vicolo dell’Osservatorio 5, 35122 Padova, Italy}

\begin{abstract}{
We present $\gtrsim 15,000$ metal-rich (${\rm [Fe/H]}>-0.2$\,dex) A and F stars whose surface abundances deviate strongly from Solar abundance ratios and cannot plausibly reflect their birth material composition. These stars are identified by their high [Ba/Fe] abundance ratios (${\rm [Ba/Fe]}>1.0$\,dex) in the LAMOST DR5 spectra analyzed by \citet{Xiang2019}. They are almost exclusively main sequence and subgiant stars with  $T_{\rm eff}\gtrsim6300$\,K. Their distribution in the Kiel diagram ($T_{\rm eff}$--$\log g$) traces a sharp border at low temperatures along a roughly fixed-mass trajectory (around $1.4\, M_\odot)$ that corresponds to an upper limit in convective envelope mass fraction of around $10^{-4}$. Most of these stars exhibit distinctly enhanced abundances of iron-peak elements (Cr, Mn, Fe, Ni) but depleted abundances of Mg and Ca. Rotational velocity measurements from GALAH DR2 show that the majority of these stars rotate slower than typical stars in an equivalent temperature range. These characteristics suggest that they are related to the so-called Am/Fm stars. Their abundance patterns are qualitatively consistent with the predictions of stellar evolution models that incorporate radiative acceleration, suggesting they are a consequence of stellar internal evolution particularly involving the competition between gravitational settling and radiative acceleration. These peculiar stars constitute 40\% of the whole population of stars with mass above 1.5\,$M_\odot$, affirming that ``peculiar" photospheric abundances due to stellar evolution effects are a ubiquitous phenomenon for these intermediate-mass stars. This large sample of Ba-enhanced chemically peculiar A/F stars with individual element abundances provides the statistics to test more stringently the mechanisms that alter the surface abundances in stars with radiative envelopes. } 
\end{abstract}

\section{Introduction} \label{Intro}
The photospheric abundances of low-mass stars largely reflect their birth material composition throughout the majority of their lifetime, over the course of stellar evolution from the main sequence (MS) to the red giant branch (RGB).
These stars have thus served as a fossil record in archaeological studies of the assembly and chemical enrichment history of our Galaxy \citep[e.g.,][]{Freeman2002, Matteucci2012, Rix2013, Ting2015, Xiang2015c, Grisoni2018, BH19, Wang2019}. On the other hand, there are a number of well-established mechanisms, both internal and external, that can alter stellar photospheric abundances to yield ``chemically peculiar stars'' outside the context of Galactic chemical evolution.  Internal atomic transport processes due to stellar evolution can both reduce the photospheric abundance of an element via gravitational setting \citep[a process that is pertinent to old, main-sequence turn-off stars;][]{Aller1960, Korn2007, Onehag2014, Choi2016, Gao2018, Souto2019} and increase the photospheric abundance of an element via radiative acceleration \citep[which may be pertinent to relatively hot stars;][]{Michaud1970, Borsenberger1984, Hui-Bon-Hoa2002, Vick2010, Michaud2011, Deal2020}.  Stellar abundances can also be altered externally, polluted by material accreted from a binary companion \citep{van_den_Heuvel1968, Boffin1988, Han1995} or via the engulfment of a planet \citep{Zuckerman2007, Church2019, Turner2019}. 

Early-type (hot) stars with exceptionally strong metal lines in their spectra have been found since the 1930's \citep[e.g.][]{Morgan1933, Titus1940, Roman1948, Cowley1969, Conti1970, Preston1974, Adelman1988, Adelman1994, Kunzli1998, Varenne1999, Adelman2007, Renson2009, Gebran2010, Royer2014, Hou2015, Monier2015, Gray2016, Ghazaryan2018, Monier2019, Qin2019}. Depending on spectral type, they were called HgMn (Mercury-manganese, $10000\lesssim T_{\rm eff}\lesssim15000$\,K) or Am/Fm (non-magnetic metal-lined A/F-type) stars \citep[e.g.,][]{Preston1974}. These stars exhibit iron-peak and heavier element abundances that are enhanced with respect to the typically solar abundances predicted for chemically `normal' stars in standard Galactic chemical evolution scenarios.  At the same time, elements, such as C, Ca and Sc, tend to be depleted in these stars. Theoretically, these peculiar abundances have been attributed to the competition between gravitational settling and radiative acceleration \citep{Michaud1970, Michaud1982, Vauclair1982, Borsenberger1984, Charbonneau1988, Alecian1996, Turcotte1998, Richer2000, Richard2001, Talon2006, Vick2010, Michaud2011, Deal2020}.     Observed abundance patterns can be reproduced reasonably well by radiative acceleration models when coupling with either rotation-induced turbulence \citep{Richer2000, Richard2001, Talon2006} or mass loss \citep{Vick2010, Michaud2011}. 

Early-type stars with shallow convective envelopes may also serve as good laboratories for studying external accretion events. This is because the accreted material mixes rapidly within the shallow convection envelope, possibly resulting in the prominent alteration of surface abundances. Although a number of external processes have been shown to yield chemically peculiar stars \citep{Havnes1971, Proffitt1989, Venn1990, Church2019}, their role in setting the peculiar abundances of Am/Fm stars in particular is not well understood.  A recent study of how planet engulfment events can lead to the formation of chemically peculiar main-sequence (turnoff) stars \citep{Church2019} further emphasizes the importance of constraining the frequency of such events.  To make advances on this issue, a large sample of early-type chemically peculiar stars with well-determined abundances for multiple elements is crucial.

At present, the Am/Fm stars in the literature that have multi-element abundance measurements are few (only a few hundred) and are mostly found in a modest number of open clusters \citep{Conti1970, Burkhart1989, Hill1993, Hill1995, Hui-Bon-Hoa1998, Varenne1999, Burkhart2000, Monier2005, Fossati2007, Gebran2008a, Gebran2008b, Gebran2010, Royer2014, Yuce2014, Monier2015, Ghazaryan2018, Monier2019}. Large-scale Galactic spectroscopic surveys implemented recently can significantly improve on this. These surveys have collected millions of stellar spectra, covering a broad range in the Hertzsprung–Russell (HR) diagram with well-defined target selection functions \citep{Carlin2012, Liu2014, De_Silva2015, Yuan2015, Majewski2017, Xiang2017b, Chen2018}. The survey spectra deliver precise abundances for more than ten elements across a large fraction of the HR diagram \citep{Ting2017, Buder2018, Xiang2019, Ahumada2019, Wheeler2020}. This makes these data sets useful not only for Galactic archaeology, but also for systematically studying chemically peculiar early-type stars in the field, which are rare in number compared to the numerous normal stars.

In this work, we report the discovery of $\sim15,000$ hot main sequence and subgiant stars with significantly or greatly enhanced s-process and iron-peak elemental abundances.  These stars have been found among the 6 million stars with low-resolution ($R\sim1800$) stellar spectra in the fifth data release (DR5) \footnote{http://dr5.lamost.org} of the LAMOST Experiment for Galactic Understanding and Exploration \citep[LEGUE;][]{Deng2012, Zhao2012}. They are distinguished by their highly enhanced Ba abundances (${\rm [Ba/Fe]}>1$\,dex), and most of them occupy only a very specific region of the $T_{\rm eff}$--$\log g$ (Kiel) diagram across the temperature range $6300\lesssim T_{\rm eff}\lesssim7500$\,K. These Ba-enhanced stars form a sharp border towards low temperatures and trace an approximately fixed stellar mass trajectory (near 1.4\,$M_\odot$). Their locations in the $T_{\rm eff}$--$\log g$ diagram and their abundance patterns arguably relates them to Am/Fm stars.  We emphasize, however, that even though Ba enhancement may be a tell-tale signature of Am/Fm stars, not all Am/Fm stars necesssarily exhibit chemical peculiarity in their Ba abundance.  

The paper is organised as follows: Section\,\ref{data} introduces the data used in this study, Section\,\ref{results} presents the results, and Section\,\ref{discussion} discusses the possible origin mechanisms of these Ba-enhanced chemically peculiar stars; Section\,\ref{conclusion} concludes.

\section{data} \label{data}
\subsection{The LAMOST and GALAH database}
We adopt the stellar abundance catalog of \citet{Xiang2019}, which includes abundance estimates for 16 elements (C, N, O, Na, Mg, Al, Si, Ca, Ti, Cr, Mn, Fe, Co, Ni, Cu, Ba) in 6 million stars derived from the LAMOST DR5 low-resolution ($R\sim1800$) spectra. Abundances are measured using {\it The DD--Payne} \citep{Ting2017, Xiang2019}, a hybrid method that combines the philosophy of data-driven spectroscopy with {\it The Payne}, a flexible spectral fitting tool based on neural network modelling \citep{Ting2019}. 
As the training set to build up the data-driven spectral model, the LAMOST {\it DD--Payne} has used the set of LAMOST stars that overlap with the GALAH \citep{De_Silva2015} and APOGEE \citep{Majewski2017} surveys (for which stellar abundance measurements exist). In the training process, priors are assigned according to the first-order derivative of Kurucz model spectra \citep{Kurucz2005} with respect to each elemental abundance, to ensure that abundances are deduced from physically sensible features in the LAMOST spectra. For a spectrum with signal-to-noise ratio (S/N) higher than 50, the statistical uncertainties on abundance estimates can be as small as 0.03\,dex for [C/Fe], [Fe/H], [Mg/Fe], [Ca/Fe], [Ti/Fe], [Cr/Fe], [Ni/Fe], and 0.15\,dex for [Ba/Fe]. Systematic errors are expected to be similar in magnitude.  In the catalogs provided by \citet{Xiang2019}, unreliable estimates are marked with a flag.   In this work, we make use of the `recommended catalog' created by \citet{Xiang2019} by combining two individual abundance catalogs scaled to either GALAH or APOGEE.  

In this paper we also draw on the GALAH DR2 \citep{Buder2018} catalog, both for validating the LAMOST results and for studying the correlation of chemical abundance peculiarities with projected stellar rotation velocity $v{\rm sin}i$. GALAH is a high-resolution ($R\sim28,000$) spectroscopic survey using the Anglo-Australian Telescope \citep{De_Silva2015, Martell2017}. Its second data release (DR2) provides stellar parameters ($T_{\rm eff}$, $\log g$, $V_{\rm mic}$, and $v\sin i$) and elemental abundances of 23 elements for 342,682 stars \citep{Buder2018}. 

\begin{figure*}[htp]
\centering
\includegraphics[width=180mm]{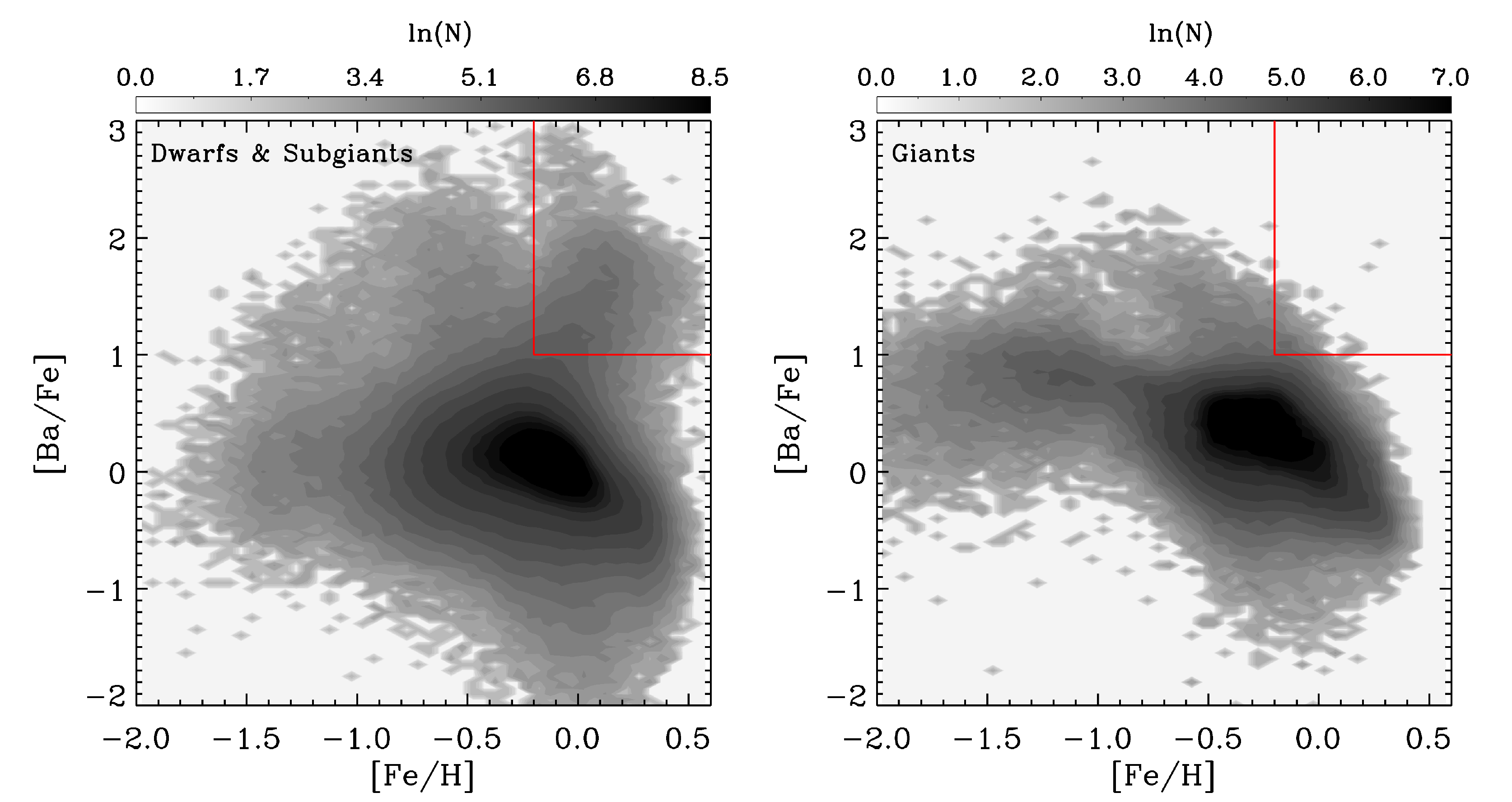}
\caption{Stellar number density in the [Ba/Fe] versus [Fe/H] plane for dwarfs and subgiants (left) and giants (right) from the LAMOST abundance catalog of \citet{Xiang2019}.  Dwarfs exhibit a prominent branch (delineated by the red lines) of metal-rich (${\rm [Fe/H]}>-0.2$\,dex) stars whose [Ba/Fe] are enhanced by 1--3 orders of magnitude with respect to solar values, whereas giants show no similar counterpart.  The majority of these stars exhibit a positive correlation between [Ba/Fe] and [Fe/H]. The stellar density is shown on a logarithmic scale.   The metal-rich, Ba-enhanced stars delineated by the red lines constitute 1\% of the entire LAMOST sample.}
\label{fig:Fig1}
\end{figure*}

The most pertinent result revealed by the \citet{Xiang2019} catalog is shown in 
Figure \ref{fig:Fig1}, which highlights where in the [Ba/Fe] versus [Fe/H] \footnote{With the conventional definition $[X/H] \equiv \log_{10}\left(\frac{N_{X}/N_{H}}{N_{X}^{\odot}/N_{H}^\odot}\right)$ for element $X$.} diagram the subset of LAMOST DR5 stars with a $g$-band spectral S/N higher than 50 are located.  This includes both dwarfs and giants, which we define as stars with $T_{\rm eff}<5600$\,K and $\log g < 3.8$ throughout this paper.  The vast majority of stars have [Ba/Fe] values around solar (${\rm [Ba/Fe]}\sim0$), but there are also a significant fraction of stars with high [Ba/Fe] (${\rm [Ba/Fe]}\gtrsim1$\,dex) or low [Ba/Fe] (${\rm [Ba/Fe]}\lesssim-1$\,dex) values.  A prominent population of metal-rich (${\rm [Fe/H]}>-0.2$\,dex), Ba-enhanced (${\rm [Ba/Fe]}>1$\,dex) dwarfs and subgiants, in particular, has no counterpart in the giant population.

\subsection{Sample selection}
We select the metal-rich and Ba-enhanced stars of interest by the criteria of ${\rm [Fe/H]}>-0.2$\,dex and ${\rm [Ba/Fe]}>1.0$\,dex, and require $S/N>50$, qflag\_chi2 = `good' to ensure that the abundance estimates are of good quality. We further discard stars classified as O-, B-type or WD according to the LAMOST pipeline \citep{Luo2015}, as their {\it DD-Payne} abundance estimates are problematic. These criteria lead to 15,009 stars in our sample. The median value of the reported measurement uncertainties for this sample of stars is 0.15\,dex in [Ba/Fe], 0.04\,dex for [Fe/H] and [X/Fe], where X = C, Mg, Si, Ca, Ti, Cr, Mn, and Ni. Our sample of Ba-enhanced stars occupy a specific region in the [Ba/Fe]--[Fe/H] plane that is populated by dwarfs and subgiants but contains almost no red giant stars. We will use the terms ``Ba-enhanced stars" and  ``Ba-enhanced chemically peculiar stars" interchangeably in this study, as we have found that these stars also exhibit chemical peculiarity in other elements.

Fig.\,\ref{fig:Fig1} demonstrates that stars can also have ``peculiar" [Ba/Fe] values when, for instance, they are Ba-enhanced (both dwarfs and giants) with ${\rm [Fe/H]}<-0.2$\,dex, or when they are metal-rich but Ba-depleted. In this study, we focus only on the metal-rich and Ba-enriched stars as defined above, which may be related to the classical Am/Fm stars. The exploration of stars with other apparent chemical ``peculiarities" is deferred to future work.

\begin{figure*}[ht!]
\centering
\includegraphics[width=180mm]{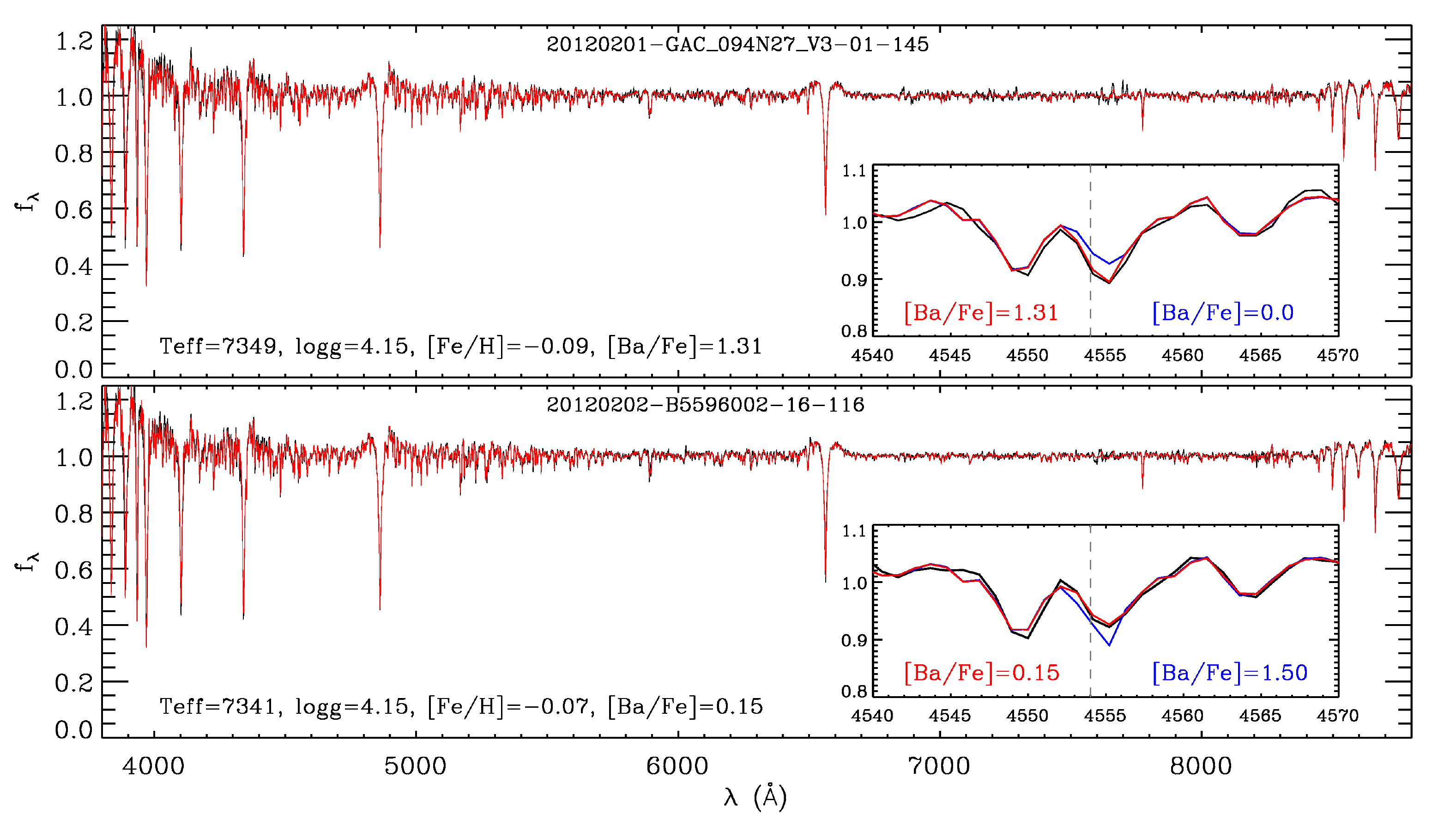}
\caption{Pseudo-continuum-normalized LAMOST spectra of a Ba-enhanced (top) and a Ba-normal (bottom) star. The black curve is the LAMOST spectrum and the red curve is the best-fitting {\it DD-Payne} model spectrum, with best-fitting parameters labeled. The zoom-in plots highlight the Ba 4554{\AA} lines. The blue curves there show the {\it DD-Payne} model spectrum with a different Ba abundance to the fitting value. Note that the pseudo-continuum is derived by smoothing the spectrum with a 50\,{\AA} wide Gaussian kernel, so that the normalized flux is not necessarily smaller than unity. }
\label{fig:Fig2}
\end{figure*}

Fig.\,\ref{fig:Fig2} shows LAMOST spectra of a Ba-enhanced A star and a Ba-normal star with similar stellar parameters. The Ba-enhanced star shows a stronger Ba\,{\sc ii} $\lambda$4554{\AA} line that cannot be fitted by the normal Ba abundance, suggesting that the Ba enhancement is a genuine feature recognizable in LAMOST low-resolution spectra. 

\begin{figure*}[htp]
\centering
\includegraphics[width=160mm]{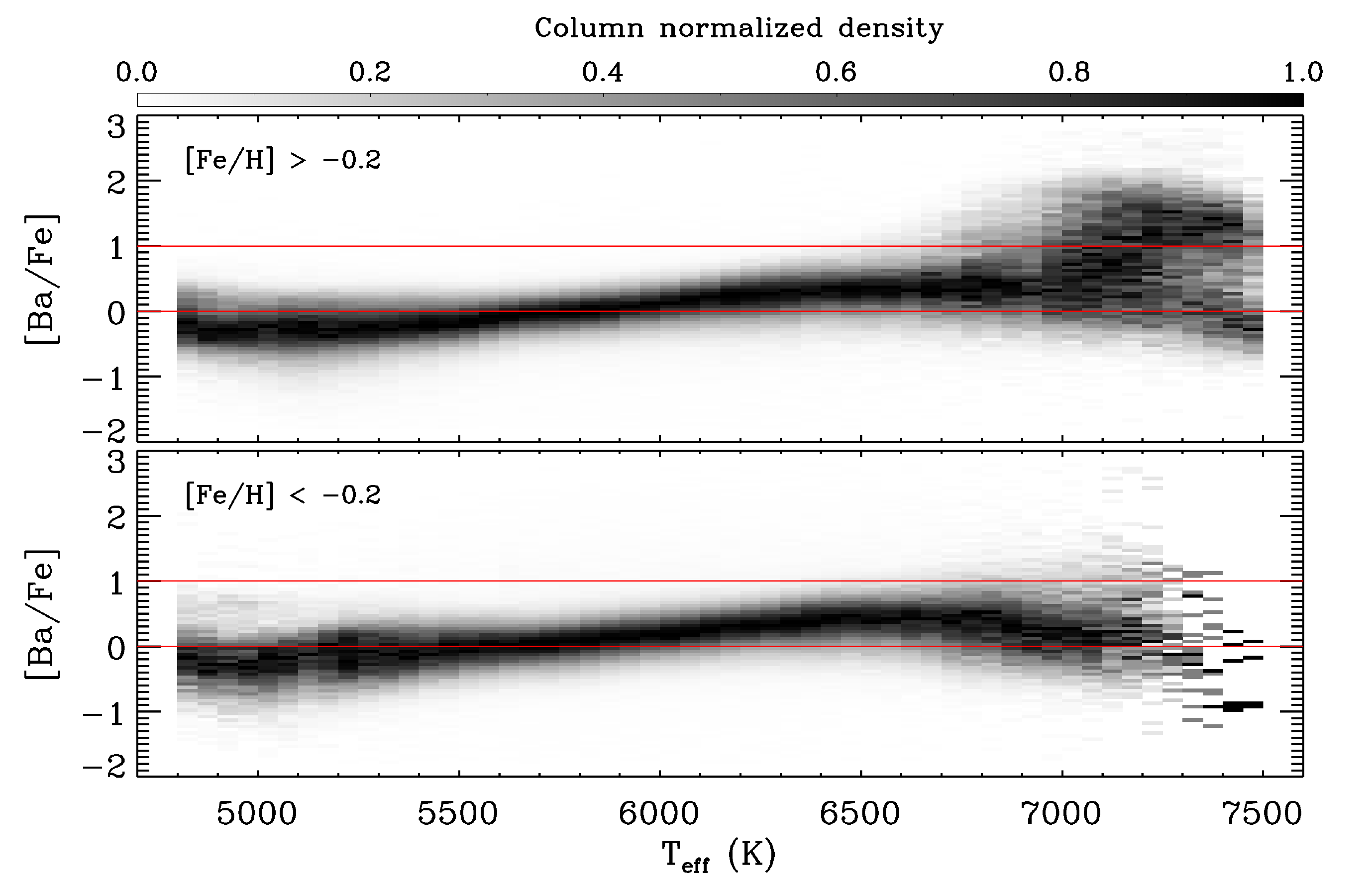}
\caption{[Ba/Fe] as a function of $T_{\rm eff}$ for dwarf and subgiant stars with ${\rm [Fe/H]}>-0.2$\,dex (top) and ${\rm [Fe/H]}<-0.2$\,dex (bottom). The color scale shows the number density of stars normalized in each column (temperature). The hot stars contain a much higher fraction of Ba-enhanced (${\rm [Ba/Fe]}>1$\,dex) stars, and a bifurcation in [Ba/Fe] is seen at the high temperature side for stars with ${\rm [Fe/H]}>-0.2$\,dex.}
\label{fig:Fig3}
\end{figure*}

Fig.\,\ref{fig:Fig3} plots [Ba/Fe] as a function of effective temperature for two subpopulations of stars, those with ${\rm [Fe/H]}>-0.2$\,dex and ${\rm [Fe/H]}<-0.2$\,dex. For both metallicity populations, there is an increasing trend in [Ba/Fe] with $T_{\rm eff}$, reaching a [Ba/Fe] value of 0.5\,dex at $T_{\rm eff}\sim6500$\,K. This trend has also been revealed by measurements from high-resolution spectra and has been argued to originate with NLTE effects \citep{Bensby2014}. Beyond 6500\,K, however, these two populations exhibit markedly different behavior. In the low metallicity ${\rm [Fe/H]}<-0.2$\,dex population, the [Ba/Fe] ratio turns over and begins decreasing with $T_{\rm eff}$, dropping back to the solar value at $T_{\rm eff}\sim7000$\,K for.  The ${\rm [Fe/H]}>-0.2$\,dex subset, on the other hand, exhibits a bifurcation in [Ba/Fe] at $T_{\rm eff}\gtrsim7000$\,K, with one branch reaching the low-[Ba/Fe] values typical of the ${\rm [Fe/H]}<-0.2$\,dex population, and a second branch of high-[Ba/Fe] values.  This latter branch constitutes our sample of Ba-enhanced stars. Such a bifurcation structure suggests that the high-[Ba/Fe] stars are not simply due to a systematic bias of the measurements, but emerge as a result of a genuine enhancement in Ba.  We verify this using high-resolution spectra in Section\,\ref{sec:highres}.     

\begin{figure}[htp]
\centering
\includegraphics[width=85mm]{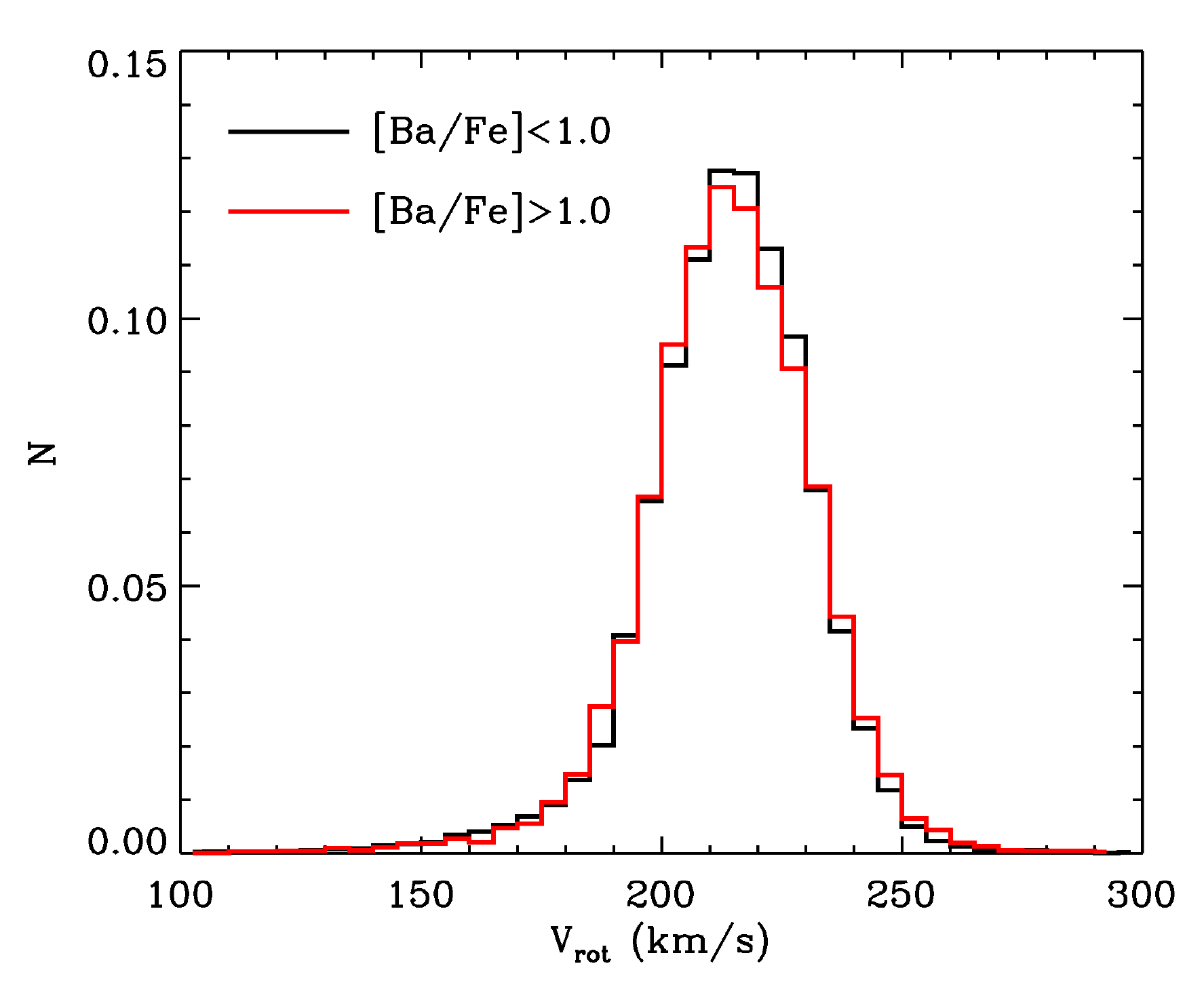}
\caption{Distribution of the Galactocentric rotation velocity for hot stars with $6700<T_{\rm eff}<7500$\,K and ${\rm [Fe/H]}>-0.2$\,dex. Stars with ${\rm [Ba/Fe]}>1$\,dex are shown in red, while stars with ${\rm [Ba/Fe]}<1$\,dex are shown in black. The  Ba-enhanced stars are perfectly typical of the Galactic thin disk.}
\label{fig:Fig4}
\end{figure}

Fig.\,\ref{fig:Fig4} highlights the distribution of Galactocentric rotational velocity for our set of Ba-enhanced stars.  The rotational velocity for each star is derived adopting the \textit{Gaia} distance from \citet{Bailer-Jones2018}, the \textit{Gaia} DR2 proper motion \citep{Brown2018} and the LAMOST DR5 radial velocity \citep[for details, see][]{Coronado2020}. The Ba-enhanced stars are indistinguishable from chemically normal metal-rich stars, and exhibit the same disk-like kinematics typical of metal-rich stars.

\subsection{Abundances verification with high-resolution spectra}\label{sec:highres}
\begin{figure*}[ht!]
\centering
\includegraphics[width=180mm]{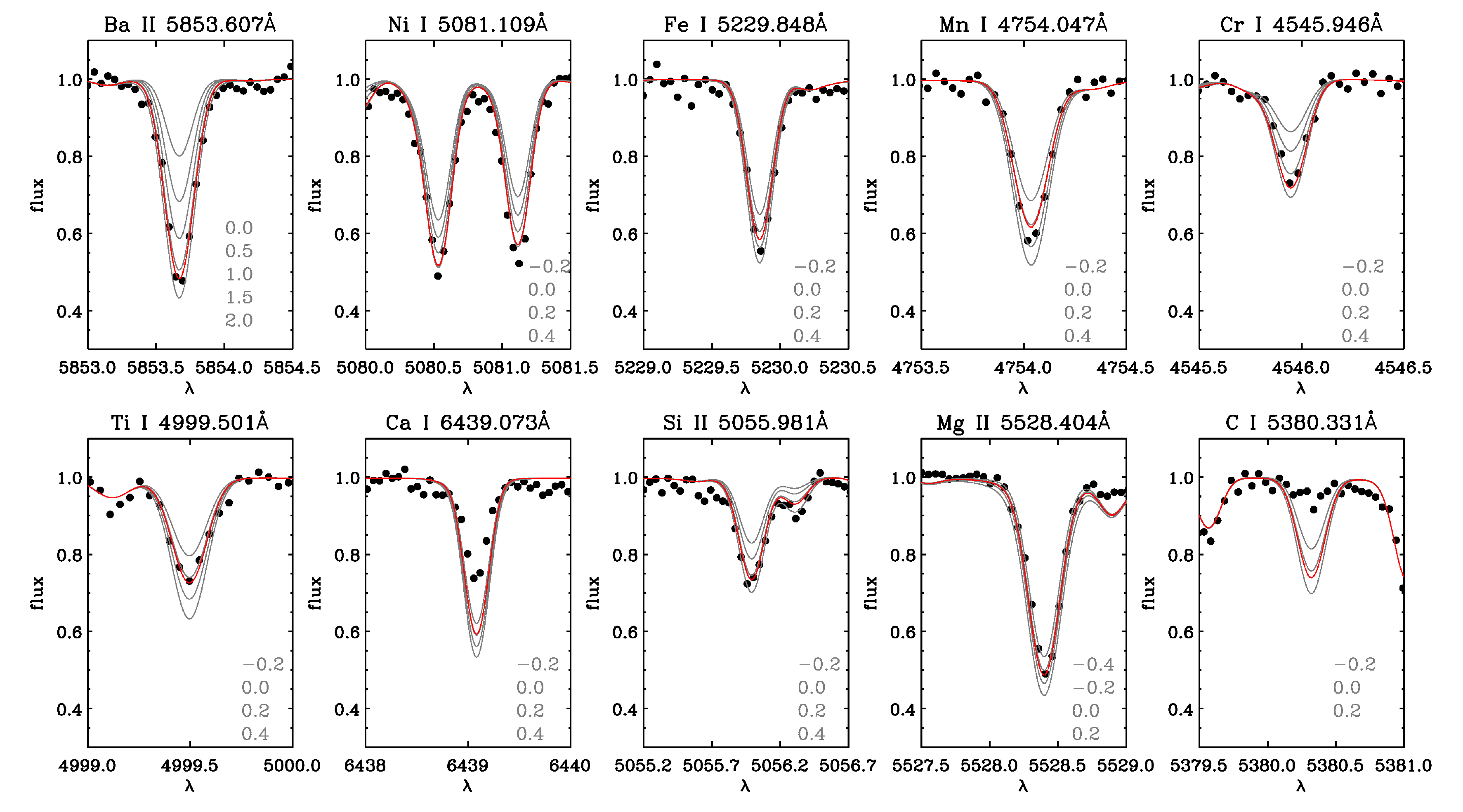}
\caption{High-resolution ($R\sim48\,000$) spectra from the FEROS spectrograph on the La Silla 2.2m telescope for an example star (Gaia ID: 3265514040885044992). The black dots are the high-resolution spectra, while solid lines are synthetic spectra from the Kurucz ATLAS12 model, adopting $T_{\rm eff} = 7375$\,K, $\log g$ = 4.31, $v\sin i$ = 3.0\,km/s, and a micro-turbulence velocity of 2.0\,km/s. Grey lines show models with different [X/H] values (marked in each panel), and the red line shows the model corresponding specifically to the [X/H] values determined with {\it The DD-Payne} applied to the LAMOST low-resolution spectra. For most elements, including Ba, the Kurucz synthetic spectra with [X/H] from the low-resolution results are consistent with the high-resolution FEROS spectra, verifying our identification of enhancement in the LAMOST spectra, and thus also our sample selection. The Ca and C lines are manifest exceptions, for reasons that we have not yet fully identified.}
\label{fig:Fig5}
\end{figure*}

The spectral signals of Ba-enhancement in the LAMOST spectra are significant, but appear subtle at face-value (Fig.\ref{fig:Fig2}). To verify them, 
we have carried out high-resolution ($R\sim48,000$) follow-up observations for a few of these Ba-enhanced stars with the FEROS spectrograph on the ESO La Silla 2.2m telescope \citep{Kaufer1999}. Figure\,\ref{fig:Fig5} shows some of the key spectral lines from the FEROS spectrum for one example star, together with the predictions from 1D-LTE Kurucz models for a range of abundances.  In all panels, we assign $v{\rm sin}i$=3.0\,km/s based on the method presented in \citet{Strassmeier1990} using the $\lambda$6432\,{\AA} line.  The Kurucz model spectra are computed adopting the $T_{\rm eff}$ and $\log g$ from the LAMOST {\it DD-Payne} catalog. The red line in each panel highlights the high-resolution spectrum predicted specifically for the abundance ratios derived from the low-resolution LAMOST spectra. In all models, the micro-turbulence velocity $V_{\rm micro}$ is set to 2.0\,km/s to match the line profile, consistent with expectations from the $V_{\rm micro}$--$T_{\rm eff}$ relation measured in the literature \citep{Bruntt2010, Gebran2014}. The models also adopt a macro-turbulence velocity $V_{\rm macro}$=8.0\,km/s, which is roughly consistent with an extrapolation of the $V_{\rm macro}$--$T_{\rm eff}$ relation in literature \citep{Bruntt2010, Doyle2014}. It should be noted, though, that this extrapolation is suited to much hotter stars than those where the $V_{\rm macro}$--$T_{\rm eff}$ relation more strictly applies (i.e. stars cooler than 6400\,K). The good agreement of line profiles between the Kurucz synthetic spectra and the high-resolution FEROS observations lends credence to this extrapolation. 

Figure\,\ref{fig:Fig5} illustrates the remarkable power of low-resolution spectra for determining even ``peculiar'' element abundances: for Ba, Ni, Fe, Mn, Cr, Ti, Si, and Mg, the observed high-resolution FEROS spectral lines are in good, though not perfect, agreement with the Kurucz model spectra (red line) adopting the abundance values derived with {\it The DD-Payne} applied to the LAMOST data.  Our focus here is on isolated single lines with good S/N but weak features to avoid saturation.  For this set of lines, NLTE effects should be mostly minimal.  For Ba, NLTE effects are expected to be negligible for the 5854{\AA} line \citep{Korotin2015}. For the Mg\,5529\,{\AA} line, the impact of NLTE effects for late A-type stars is smaller than 0.1\,dex \citep{Alexeeva2018}. In the case of Cr, Mn, Fe and Ni, the spectral lines are even stronger features than in the Kurucz model that adopts the best fitted {\it DD-Payne} abundance, which is opposite to the expected impact of NLTE effects and possibly implying that the star is iron-peak enhanced. The Ca and C~{\sc i} lines are glaring exceptions that we will follow-up with more thorough analysis in the future. We emphasize, though, that the chemical peculiarity in these cases is unlikely due to NLTE, given that this should be ubiquitous for all stars at this temperature and $\log g$. 

\section{Results} \label{results}
On the basis of the consistency between the low-resolution LAMOST spectra and our high resolution FEROS data, {\it DD-Payne} abundances derived from LAMOST spectra should yield a representative sample of genuinely Ba-enhanced stars.  Here we examine the distribution of these Ba-enhanced stars in the Kiel-diagram and in $v{\rm sin}i$.  Our goal is to characterize their detailed pattern of abundance peculiarity and their overall incidence among stars of the same temperature.

\subsection{Distribution in the $T_{\rm eff}$--$\log g$ diagram}
\begin{figure}[ht!]
\centering
\includegraphics[width=85mm]{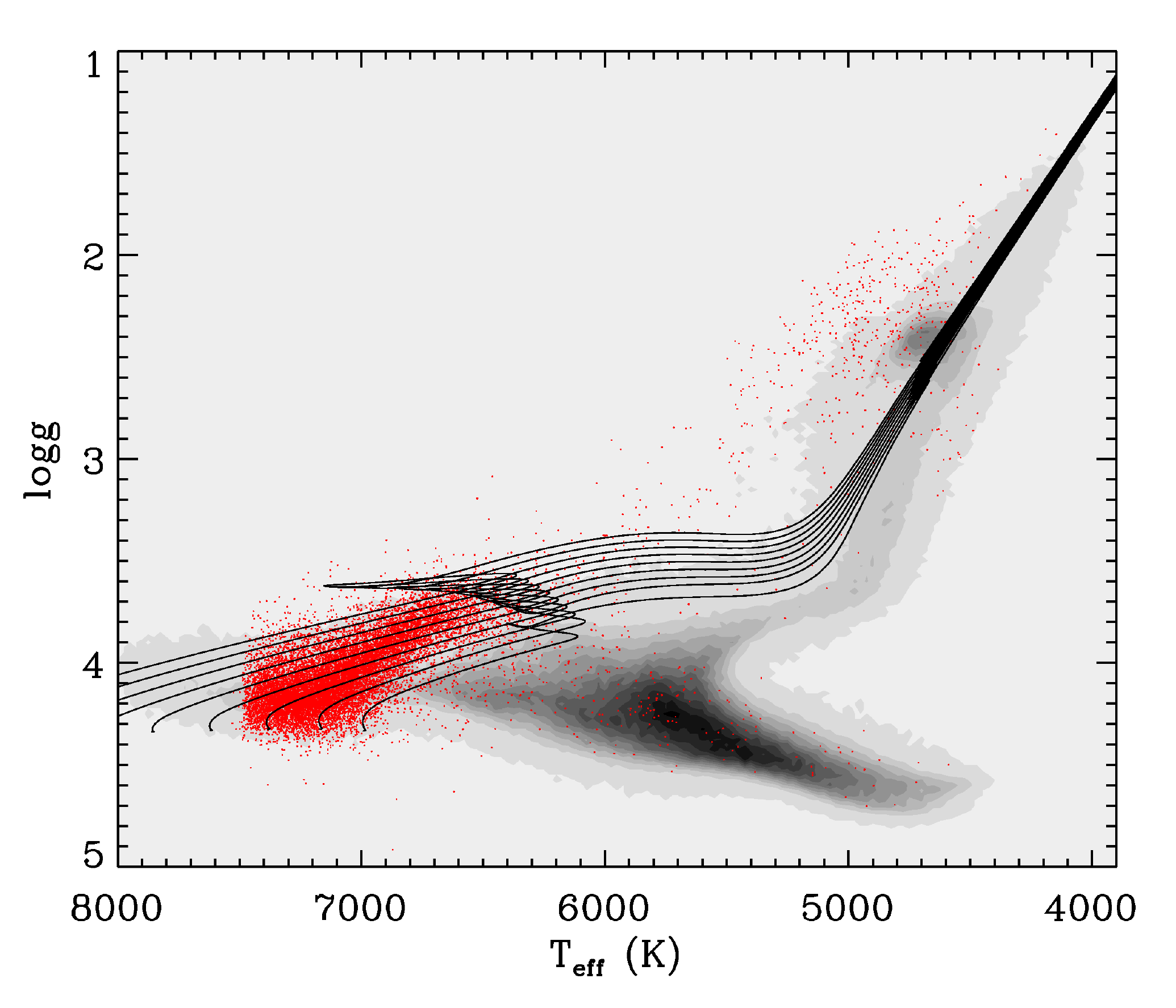}
\caption{Distribution of the sample of Ba-enhanced stars (red) in the $T_{\rm eff}$ -- $\log g$ diagram. The color-coded background shows the number density of all stars with ${\rm [Fe/H]}>-0.2$\,dex. There is a hard cut among Ba-enhanced stars at 7500\,K, above which the {\it DD-Payne} abundances are not available in the recommended catalog of \citet{Xiang2019}. PARSEC stellar evolution tracks with solar metallicity and with mass uniformly distributed from 1.4 to 1.8\,$M_\odot$ are overplotted. } 
\label{fig:Fig6}
\end{figure}

 The distribution of the metal-rich, Ba-enhanced stars in the $T_{\rm eff}$--$\log g$ (Kiel) diagram is plotted in Fig.\,\ref{fig:Fig6}. From their location in the plot, we can see that these stars are almost exclusively main-sequence (turnoff) and subgiant stars with relatively high temperature ($T_{\rm eff}\gtrsim6300$\,K). Their distribution exhibits a sharp border towards low $T_{\rm eff}$ (aside from the imposed temperature cut at $T_{\rm eff}<7500$~K), with few, if any, stars at the lowest temperatures. The lack of cooler stars immediately implies that the origin of these Ba-enhanced stars cannot lie in normal Galactic chemical evolution, which would predict stars across a broad temperature range, including at the low $T_{\rm eff}$ side of the diagram. 

The shape of the low-temperature border -- which traces a decrease in $T_{\rm eff}$ with decreasing $\log g$ -- can yield insight into the nature of these stars.  We study this further in Fig.\,\ref{fig:Fig6}, which incorporates the PARSEC stellar evolution tracks \citep{Bressan2012}.
The distribution of Ba-enhanced stars in the Kiel diagram is consistent with the main-sequence (turnoff) and subgiant regimes of the stellar evolutionary tracks for (initial) stellar mass between 1.4 and 1.8\,$M_\odot$. The lower border is consistent with a mass in the range 1.4--1.5\,$M_\odot$, depending on $\log g$.

\begin{figure}[htp]
\centering
\includegraphics[width=85mm]{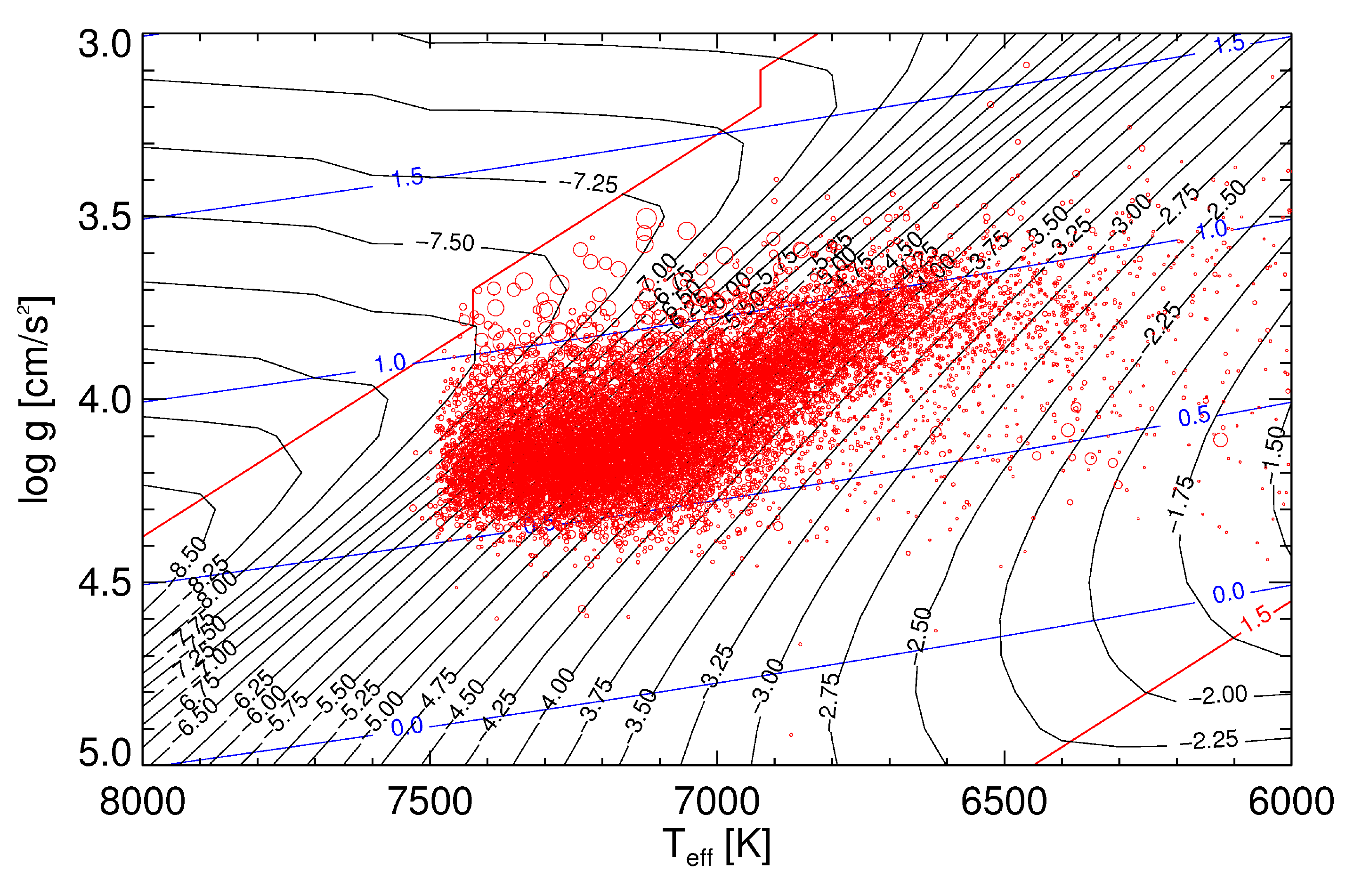}
\caption{Convective envelope to stellar mass ratio log($M_{\rm cenv}$/$M_*$) from the stellar envelope model of \citet{Ludwig1999}. Red circles mark the observed Ba-enhanced stars, with the size of the circle representing the value of [Ba/Fe]. The red line at the upper left marks the location where the convective envelopes split into two separate zones. The area to the lower right of the red line at the bottom-right corner is the region in which the convective envelopes are not fully embedded in the computational domain anymore. The blue lines depict the estimated luminosity log($L$/$L_\odot$).}
\label{fig:Fig7}
\end{figure}

In Fig.\,\ref{fig:Fig7} we compare the distribution of our sample of Ba-enhanced stars with the stellar envelope models from \citet{Ludwig1999} for a range of convective envelope mass ratios in the $T_{\rm eff}$ -- $\log g$ plane.  The low-$T_{\rm eff}$ border of these Ba-enhanced stars corresponds to a convective envelope mass ratio of about $\simeq10^{-4}$, and this convective envelope mass fraction drops by four orders of magnitude towards the hottest stars. This offers a satisfying stellar evolutionary explanation for the distribution of Ba-enhanced stars and provides a valuable diagnostic of external accretion events (see Section\,\ref{discussion}).

Finally, we note that $\sim$3\% (482) of our sample are giants, with $T_{\rm eff}<5600$\,K and $\log g<3.6$. Among these giants, only 12 of them (or 2.5\%) have ${\rm [Ba/Fe]>1.5}$\,dex, a much lower fraction than that of the overall sample ($\sim 1/3$). Throughout this work, we will focus on the relatively hot A/F stars. The nature of the cooler Ba-enhanced stars will be explored in an upcoming study \citep[Zhang et al. in prep., see also][]{Norfolk2019}. 

\subsection{Distribution in abundance space} \label{abundancepattern}
Although {\it The DD-Payne} is designed to provide abundances for 16 elements from the LAMOST spectra, for the relatively hot stars of interest in this work, some abundances can not be determined \citep[i.e. given spectral features too weak to allow for reliable abundance determination;][]{Xiang2019}, and the fraction of stars with useful abundance estimates varies. 

\begin{figure*}[htp]
\centering
\includegraphics[width=180mm]{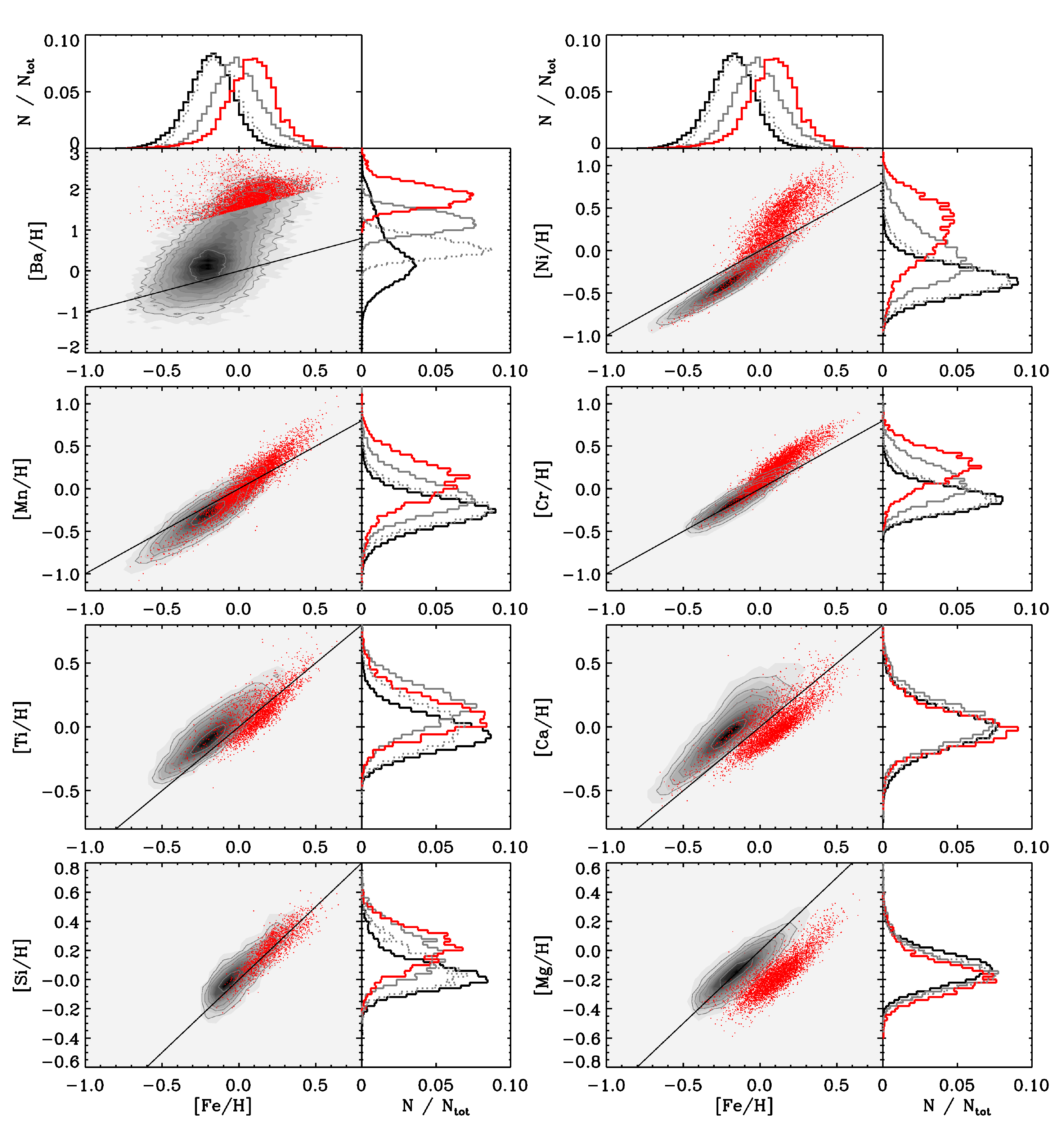}
\caption{[X/H] versus [Fe/H] plane for stars with $6700<T_{\rm eff}<7500$\,K. The background contour shows the number density of stars with ${\rm [Ba/Fe]}<0.5$\,dex except for the panel for [Ba/H], in which the background contour shows the number density of all stars. The red dots show the stars with ${\rm [Ba/Fe]}>1.5$\,dex. The histograms in the right show the [X/H] distribution of stars with ${\rm [Ba/Fe]}<0.5$\,dex (black), $0.5<{\rm [Ba/Fe]}<1.0$\,dex (grey dotted), $1.0<{\rm [Ba/Fe]}<1.5$\,dex (grey solid), and ${\rm [Ba/Fe]}>1.5$\,dex (red), respectively. The most Ba-enhanced stars are also enhanced in iron-peak elements (Cr, Mn, Fe, Ni), Ti and Si, but not enhanced in Ca, and even depleted in Mg.}
\label{fig:Fig8}
\end{figure*}

Fig.\,\ref{fig:Fig8} highlights the Ba-enhanced sample in elemental abundance space [X/H]--[Fe/H] for X = Mg, Si, Ca, Ti, Cr, Mn, Fe, Ni, and Ba for stars with $6700<T_{\rm eff}<7500$\,K. Ba-enhancement coincides with an enhancement of iron-peak elements (Cr, Mn, Fe, Ni), as well as with Si and Ti. Mg and Ca, on the other hand, show no significant overabundance. Mg even exhibits a slight underabundance. For the most Ba-enhanced (${\rm [Ba/Fe]}>1.5$\,dex) stars, [Fe/H] is 0.3\,dex higher than that of the Ba-normal stars on average.  For this set, the overabundance of Cr and Ni are even more prominent, reaching about 0.5\,dex and 0.8\,dex, respectively. Ba can be enhanced by up to 1000 times (3\,dex), making this the most overabundant element.

Wide binaries are ideal sources to verify this pattern of abundances.  The member stars in a wide binary system should have identical initial elemental abundances, given that they are likely to have formed in the same birth cloud.  Wide binary pair stars have the added advantage that they can be targeted by LAMOST due to their wide separation. \citet{El-Badry2018} identified more than 50,000 wide binary systems using parallax and proper motions in \textit{Gaia} DR2. Among them, 4714 systems have LAMOST DR5 spectra for both of the pair stars. About 80\% of these systems have a physical separation smaller than 0.08\,pc, and the maximal separation is 0.23\,pc. Such small separations suggest they are indeed co-natural \citep[e.g.][]{Kamdar2019}. From this set of stars we select systems with ${\rm S/N}>40$ for both stars in the pair, and additionally require that the pair star with the lower temperature has $T_{\rm eff}<6500$\,K, in an effort to insure that these stars have chemically normal abundances. This yields a total of 1200 wide binary pairs that can be analyzed in the manner described above.  Of these, 15 pairs hold a Ba-enhanced member with ${\rm [Ba/Fe]}>1$\,dex in the temperature range of $6700<T_{\rm eff}<7500$\,K.  
  
\begin{figure*}[ht!]
\centering
\includegraphics[width=160mm]{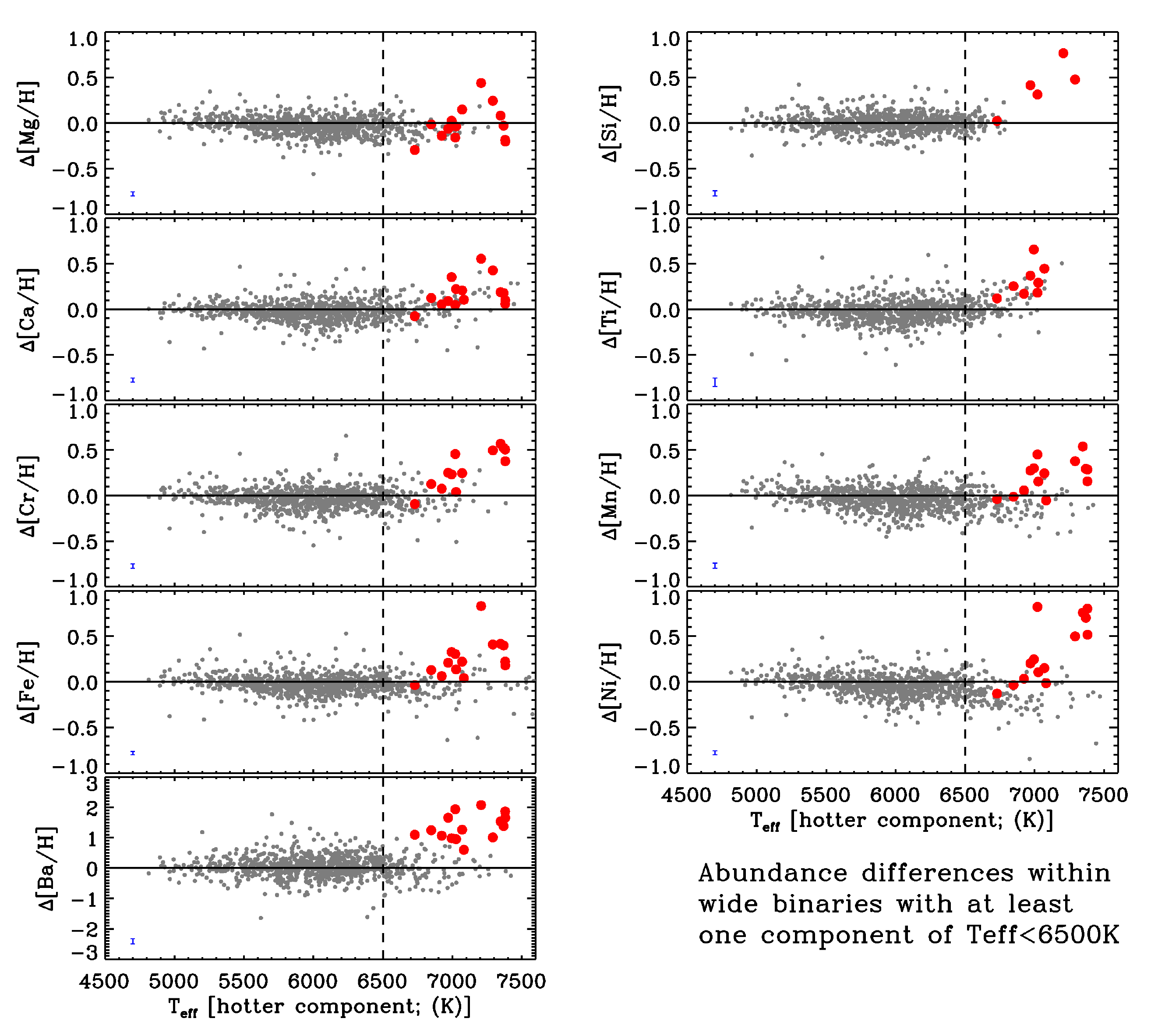}
\caption{Abundance differences between wide binary pair stars as a function of $T_{\rm eff}$. The vertical axis shows abundances for the higher-$T_{\rm eff}$ pair star minus those of the lower-$T_{\rm eff}$ pair star. The horizontal axis shows the $T_{\rm eff}$ of the higher-$T_{\rm eff}$ pair star. 
Red dots are chemical anomalies with ${\rm [Ba/Fe]}>1.0$\,dex for the higher-$T_{\rm eff}$ star of the pair. All the lower-$T_{\rm eff}$ pair stars have temperatures of $4500<T_{\rm eff}<6500$K. The typical measurement error on [X/H] is shown at the bottom-left corner of each panel and is $\sim$0.05\,dex for Mg, Si, Ca, Cr, Mn, Fe, and Ni, and $\sim$0.1\,dex for Ti, and $\sim$0.15\,dex for Ba. Note that there are fewer stars with available Si and Ti abundance measurements than all other elements.}
\label{fig:Fig9}
\end{figure*}

Fig.\,\ref{fig:Fig9} shows that the abundances of iron-peak elements (Cr, Fe, Mn, Ni) in the set of Ba-enhanced pair stars are enhanced compared to their low-$T_{\rm eff}$ companions, whereas stars with ${\rm [Ba/Fe]}<1$\,dex in the same temperature range exhibit little to no enhancement with respect to their low-$T_{\rm eff}$ companions. 
On the other hand, the [Mg/H] for the majority of the Ba-enhanced stars are slightly underabundant with respect to their low-$T_{\rm eff}$ companions.  This is similar to our finding for the overall sample.  Meanwhile, Si and Ti abundances for the Ba-enhanced stars are moderately enhanced compared to their low-$T_{\rm eff}$ companions, again consistent with findings for the overall sample. 

\subsection{Incidence of the Ba-enhanced A/F stars}
\begin{figure*}[ht!]
\centering
\includegraphics[width=180mm]{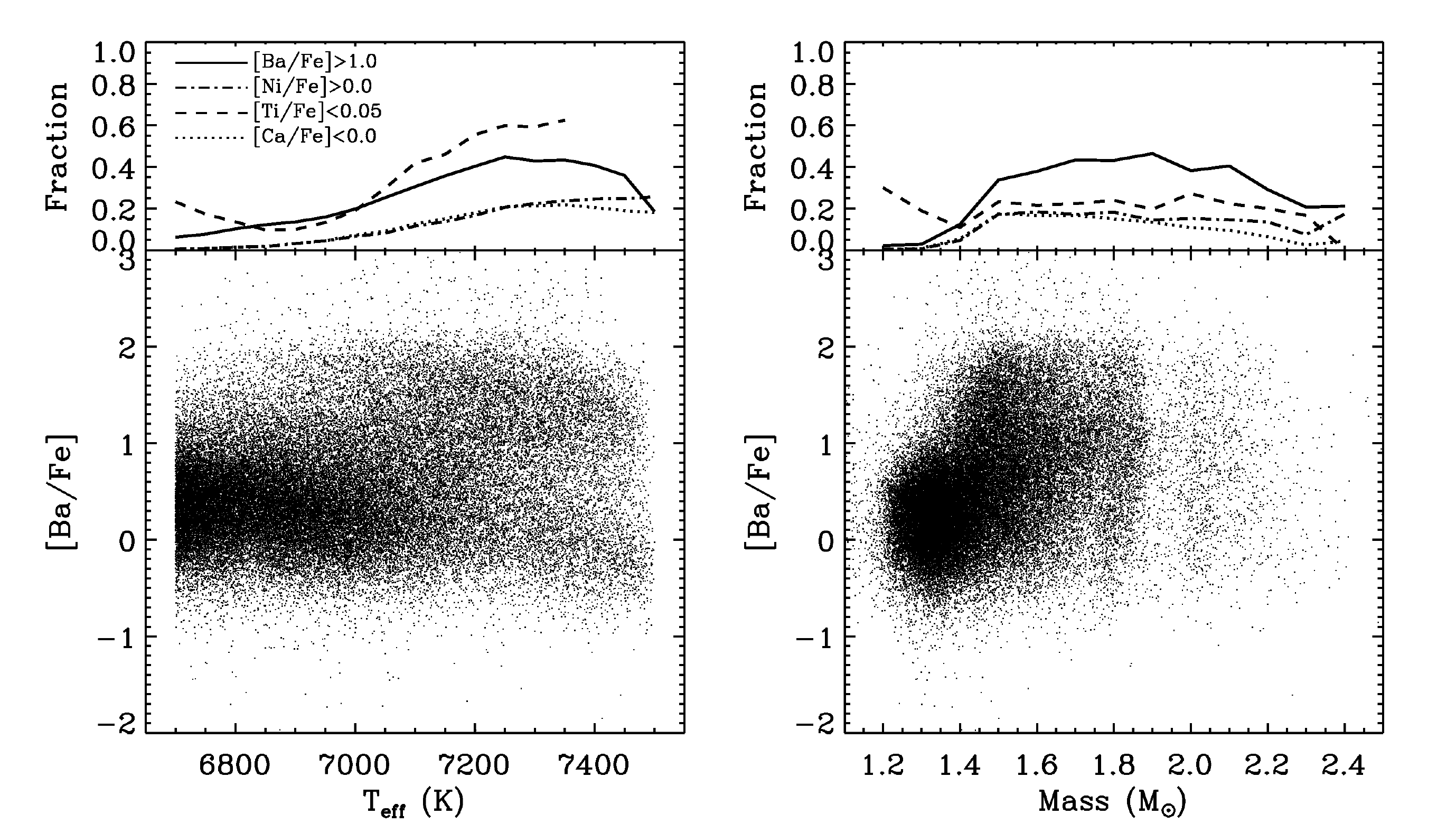}
\caption{Fraction of chemically peculiar stars as a function of effective temperature (left) and stellar mass (right).}
\label{fig:Fig10}
\end{figure*}
The above metal-rich, Ba-enhanced chemically peculiar stars constitute 16\% of the entire stellar population within the temperature range $6700<T_{\rm eff}<7500$\,K. Fig.\,\ref{fig:Fig10} illustrates an increase in the fraction of Ba-enhanced stars with both effective temperature and stellar mass. The fraction could be as high as 40\% at temperatures above 7200\,K or stellar masses higher than 1.5\,$M_\odot$. Note that we see a slightly decreasing trend at the high-$T_{\rm eff}$ and high-mass ends. Given the imposed hard cut at 7500\,K in the abundance estimates derived with {\it DD-Payne} and the small number of high-mass ($>2$\,$M_\odot$) stars, it is difficult to unambiguously relate such a trend to either a genuine phenomenon or to data artifacts. 
The trend can be also sensitive to the mass estimates used here, which is difficult to assign for chemically peculiar stars, given that stellar models are not precise enough to account for the atomic diffusion effects experienced by these stars.  Our mass estimates are derived by matching stellar isochrones from the Dartmouth Stellar Evolution Database \citep[DSEP;][]{Dotter2008} using the Bayesian method presented in \citet{Xiang2019}, assuming that all stars have an initial metallicity of [Fe/H] = $-0.1$\,dex and [$\alpha$/Fe] = 0.0\,dex.  

Fig.\,\ref{fig:Fig10} also shows the fraction of peculiar stars defined with respect to other elements, selected according to the abundance pattern revealed in Fig.\,\ref{fig:Fig8}. About 10--20\% of the stars with $T_{\rm eff}>7000$\,K are found to have enhanced nickel abundances with respected to iron (${\rm [Ni/Fe]}>0$), and a similar fraction of stars have depleted calcium abundances with respect to iron (${\rm [Ca/Fe]}<0$).   
  
The high incidence of these Ba-enhanced stars, alongside their distribution in $T_{\rm eff}$--$\log g$ space and elemental abundance patterns, suggests that these stars are plausibly related to (but not necessarily identical with) the previously known Am stars \citep{Conti1970, Smith1971, Preston1974, Abt1981, Gray2016, Qin2019}. In particular, \citet{Gray2016} identified 1067 Am stars from the LAMOST spectral database in the $Kepler$ field utilizing the MK classification method, corresponding to an Am frequency of 34.6\% for stars with spectral type between A4 and F1. In a dedicated search of metal-line stars in LAMOST DR5, \citet{Qin2019} identified 10,503 Am/Fm stars with the random forest method and reported an incidence of about 22\% for stars with $(J-H) > 0.1$. 
These numbers are qualitatively consistent with ours, with slight differences possibly due to differences in the approach used for peculiar star characterization.  On closer inspection, however, we find that the methods do not always select identical sets of stars. The cross-match of our sample with \citet{Gray2016} using the LAMOST spectra ID yields only 258 stars in common, 175 of which are identified as either Am stars or stars with strong Sr lines.  A cross-match with \citet{Qin2019} yields only 1881 stars in common. This relatively low overlap is in part due to the conservative design of {\it The DD-Payne}, which recommends Ba abundance determinations only when they can be deemed physically sensible \citep{Xiang2019}. On the other hand, it is known that some Am/Fm stars do not exhibit Ba-enhancement \citep{Ghazaryan2018}.  Thus, our selection of Ba-enhanced stars omits a fraction of the Am/Fm population. We therefore emphasize that, although we conclude from this study that our sample is mostly related to Am/Fm stars, we do not necessarily expect it to be a strict subset of the Am/Fm stars identified by \citet{Qin2019}. 
  
\subsection{$v{\rm sin}i$ from GALAH DR2}
\begin{figure*}[htp]
\centering
\includegraphics[width=180mm]{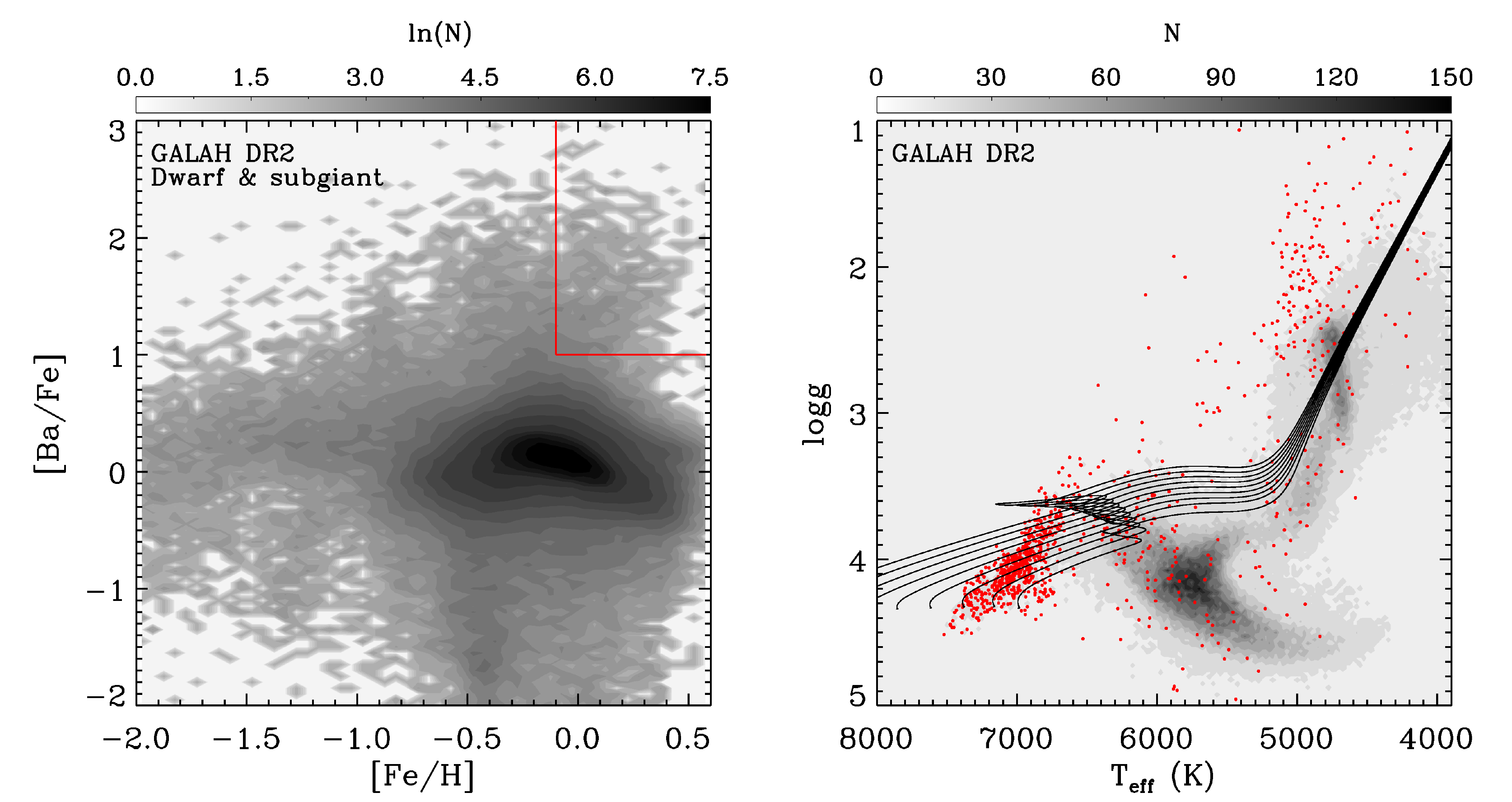}
\caption{{\em Left}: The [Ba/Fe] versus [Fe/H] diagram of GALAH DR2 dwarf and subgiant stars. {\em Right}: $T_{\rm eff}$ -- $\log g$ diagram of the GALAH DR2 stars with ${\rm [Fe/H]}>-0.1$\,dex. The color-coded background shows the number density of stars. Red points are stars with ${\rm [Ba/Fe]}>1$\,dex, as delineated by the boundary lines in red in the left panel. The PARSEC stellar evolution tracks with solar metallicity and with mass uniformly distributed from 1.4 to 1.8\,$M_\odot$ are also shown.}
\label{fig:Fig11}
\end{figure*}

Am/Fm stars have been suggested to be mostly slow rotators compared to chemically normal stars \citep{Abt1995, Abt2000}. Considering that the resolution of LAMOST spectra is too low to yield robust measurements of rotation velocity, here we make use of the GALAH DR2 catalog \citep{Buder2018} to examine the rotation of Ba-enhanced A/F stars. Fig.\,\ref{fig:Fig11} shows [Ba/Fe] vs. [Fe/H] as well as the $T_{\rm eff}$--$\log g$ diagram from GALAH DR2. We have selected stars with good S/N in the GALAH spectra by requiring $S/N>20$ for both the $c2$ and $c3$ spectral segments. No cut based on abundance quality flags in GALAH DR2 is adopted since this would discard the peculiar stars that are of interest here; we find that all stars with ${\rm [Ba/Fe]}>1.0$\,dex are flagged (i.e., marked as questionable estimates) in GALAH DR2. This flagging indicates that the abundance measurements are extrapolations outside the range of the training set.  But as we have demonstrated above, the spectral features are strong, suggesting that the measurements still have value in a relative sense.  Indeed, in the distribution of selected stars across the $T_{\rm eff}$--$\log g$ diagram in Fig.\,\ref{fig:Fig11} we see the same behavior as exhibited by the LAMOST sample in Fig.\,\ref{fig:Fig6}. Note that here we show stars with ${\rm [Ba/Fe]}>1.0$\,dex and ${\rm [Fe/H]}>-0.1$\,dex, rather than ${\rm [Fe/H]}>-0.2$\,dex as adopted for the LAMOST data, given that [Fe/H] of GALAH DR2 was found to be 0.05--0.1\,dex higher than the recommended value in the LAMOST catalog based on the APOGEE (Payne) scale \citep{Xiang2019}.  

\begin{figure*}[htp]
\centering
\includegraphics[width=180mm]{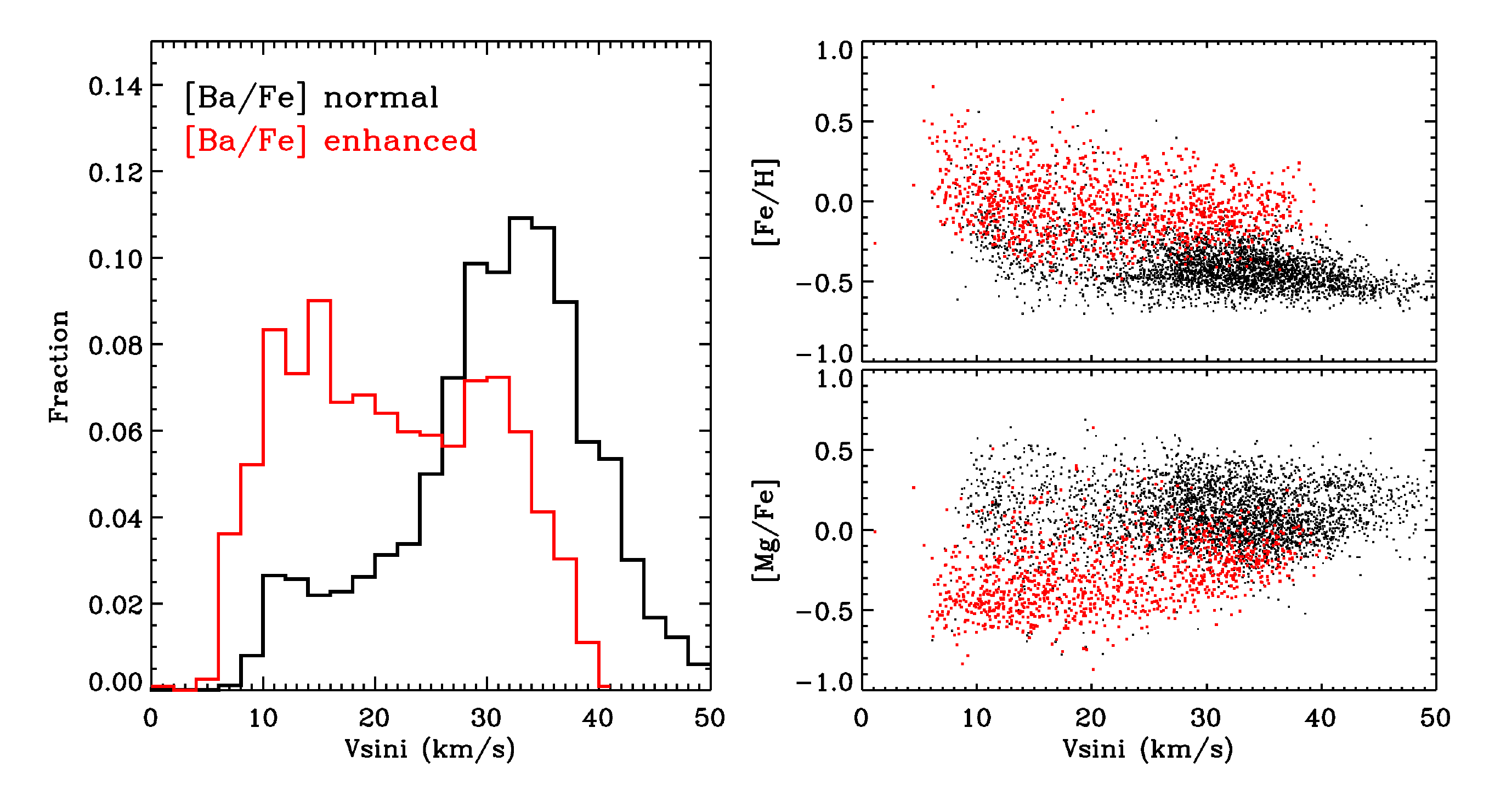}
\caption{{\em Left:} GALAH DR2 $v{\rm sin}i$ distributions for Ba-normal (black) and Ba-enhanced (red) stars. {\em Right:} [Fe/H] (top) and [Mg/Fe] (bottom) as a function of $v{\rm sin}i$ for Ba-normal (black) and Ba-enhanced (red) stars.}
\label{fig:Fig12}
\end{figure*}

Considering the significant variation of [Ba/Fe] from GALAH DR2 with $T_{\rm eff}$, we introduce a further $T_{\rm eff}$-dependent criterion to differentiate between Ba-normal and Ba-enhanced stars, thereby yielding a clean sample that can be used to characterize the shape of the Ba-enhanced $v{\rm sin}i$ distribution. Specifically, we compute the median and dispersion of [Ba/Fe] in 100\,K wide $T_{\rm eff}$ bins and the take stars with [Ba/Fe] values deviating by more than 3$\sigma$ from the median as the sample of Ba-enhanced stars. In total, there are 1188 Ba-enhanced stars with $6700<T_{\rm eff}<7500$\,K, constituting 18\% of the total number of stars in this temperature range, as is consistent with the LAMOST results (16\%).  The $v{\rm sin}i$ distribution of these Ba-enhanced stars from GALAH DR2 are shown in Fig.\,\ref{fig:Fig12}.  The majority of the Ba-enhanced stars rotate slower than the normal stars, consistent with the Am stars studied by \citet{Abt2000}. 

However, a considerable fraction of Ba-enhanced stars have large $v{\rm sin}i$; around 37\% have $v{\rm sin}i$$>$ 25\,km/s, which is comparable to the bulk of the normal stars. These stars may have slightly different levels of abundance enhancement than the remainder of the Ba-enhanced population. As we found in Section\,\ref{abundancepattern} with the LAMOST data, Ba-enhanced stars exhibit enhanced [Fe/H] but depleted [Mg/Fe] with respect to the normal stars on the whole. However, the Ba-enhanced stars with larger $v{\rm sin}i$ values tend to have slightly larger [Mg/Fe] than the slower rotators. On the other hand, 18\% of the Ba-normal stars have a $v{\rm sin}i$ smaller than 20\,km/s. These stars tend to have higher [Fe/H] than the stars with larger $v{\rm sin}i$, indicating they are probably also peculiar to some extent.  It should be noted, though, that their [Mg/Fe] values are comparable to the faster rotators and systematically higher than the Ba-enhanced stars. 

As a final note, we highlight that the GALAH DR2 $v{\rm sin}i$ distribution for the Ba-normal stars exhibits a peak near $\sim35$\,km/s. This is systematically lower than found in the literature \citep[e.g.][]{Abt2000}; previous studies have suggested that a large fraction of A stars can be expected to have $v{\rm sin}i$ larger than 100\,km/s. This discrepancy can mainly be attributed to the tendency for the underestimation of $v{\rm sin}i$ at the high-$v{\rm sin}i$ end by GALAH DR2, which will be improved on by GALAH DR3 (S. Buder, private communication).  Given that GALAH DR2 $v{\rm sin}i$ estimates are robust in the relative sense, however, this does not affect the main conclusion of this paper. 
    
\section{discussion} \label{discussion}
In this section we discuss the origin of Ba-enhanced chemically peculiar stars, focusing on one of the most likely internal mechanisms, stellar radiative acceleration.  We also consider possible external mechanisms, such as the accretion of materials from companion stars or planets.
\subsection{Stellar radiative acceleration} \label{radiativeacceleration}
The properties presented in Section\,\ref{results} imply that the Ba-enhanced chemically peculiar A/F stars are related to the Am/Fm stars that have been widely studied in the literature. Extensive effort has been made to describe these stars in the context of atomic diffusion due to stellar radiative acceleration \citep{Michaud1970, Turcotte1998, Richer2000, Richard2001, Talon2006, Vick2010, Michaud2011, Deal2020}, and the abundance patterns of elements from C to Ni have indeed been reproduced quite well by introducing mixing mechanisms such as turbulence \citep{Turcotte1998, Richer2000, Richard2001} and mass loss / stellar winds \citep{Vick2010, Michaud2011}. 
   
\begin{figure}[ht!]
\centering
\includegraphics[width=85mm]{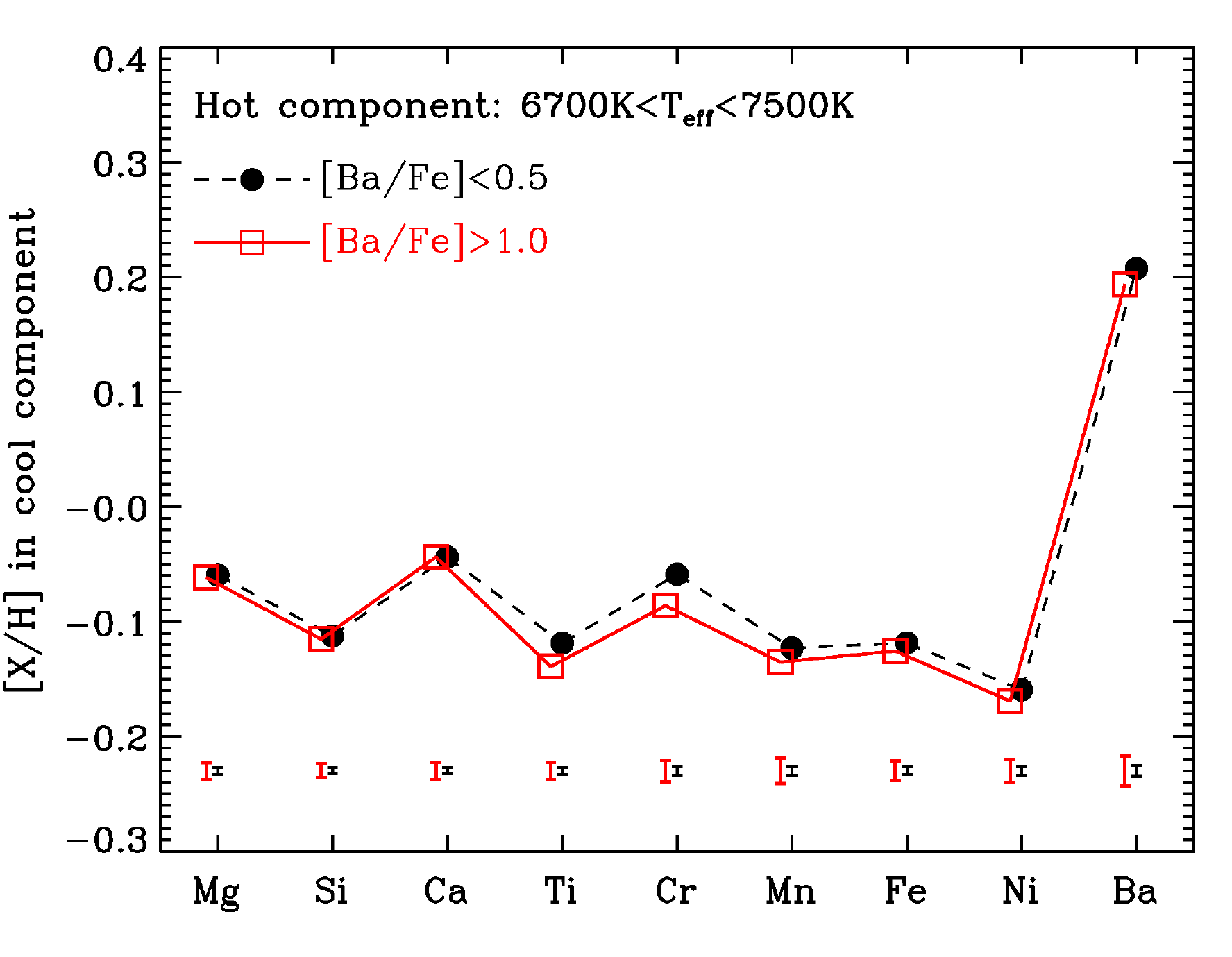}
\caption{Mean abundances of the low-$T_{\rm eff}$ ($T_{\rm eff}<6500$\,K) stars in wide binary pairs with either Ba-normal (black) or Ba-enhanced (red) high-$T_{\rm eff}$ companion stars with $6700<T_{\rm eff}<7500$\,K.  Error bars on the mean abundances are shown at the bottom of the figure. The abundance patterns among the low-$T_{\rm eff}$ companions of chemically peculiar and normal stars are indistinguishable, as expected (see text).  Differences among the hotter companion stars thus reflect stellar evolution effects rather than compositional differences in birth material.}
\label{fig:Fig13}
\end{figure}

Here we compare our observed abundance patterns with the stellar radiative acceleration models from the literature, making two main assumptions. Our first main assumption is that, on average, the Ba-enhanced chemically peculiar stars share the same initial (birth) abundances as chemically normal stars of similar mass. This is supported by several direct and indirect pieces of evidence. Firstly, the similarity in their spatial distributions and kinematics, and the fact that all main sequence and subgiant stars with masses larger than 1.4\,$M_\odot$ are younger than $\sim$2.5\,Gyr, suggests that the birth metallicities of Ba-enhanced chemical peculiars should not be very different from that of normal stars. Indeed, in the temperature range $6700<T_{\rm eff}<7500$\,K, stars more massive than 1.5\,$M_\odot$ are subgiants (Fig.\,\ref{fig:Fig6}), for which there is a tight relation between mass and age (where age mostly indicates main-sequence lifetime), implying that stars with similar mass also have similar age. Meanwhile, stars with similar ages can be expected to have similar birth metallicities, given that the dispersion in the metallicity of the interstellar medium at a given radius in disk galaxies is small \citep[$<0.1$\,dex, e.g.,][]{Kreckel2019}.  Secondly, and more directly, the low-$T_{\rm eff}$ star in \textit{Gaia} wide binary pairs have identical abundances when the pair star is either a Ba-enhanced chemically peculiar star or a normal star (Fig.\,\ref{fig:Fig13}). The abundance of the low-$T_{\rm eff}$ pair star in a \textit{Gaia} wide binary system serves as an indicator of the birth abundance of the binary system (and thus the high-$T_{\rm eff}$ star) since neither stellar evolution processes nor external accretion events will have a large impact given the thick convective envelopes and young ages of these stars. 

Our second main assumption is that chemically normal stars do not experience the radiative acceleration process or, if they do, that the effects are negligible compared to impact on Ba-enhanced chemically peculiar stars.  In this way, differential abundances between peculiar stars and normal stars can be taken as indicators of the impact of the radiative acceleration process on elemental abundance enhancements. We emphasize that the validity of this assumption is in need of verification.  

\begin{figure*}[ht!]
\centering
\includegraphics[width=180mm]{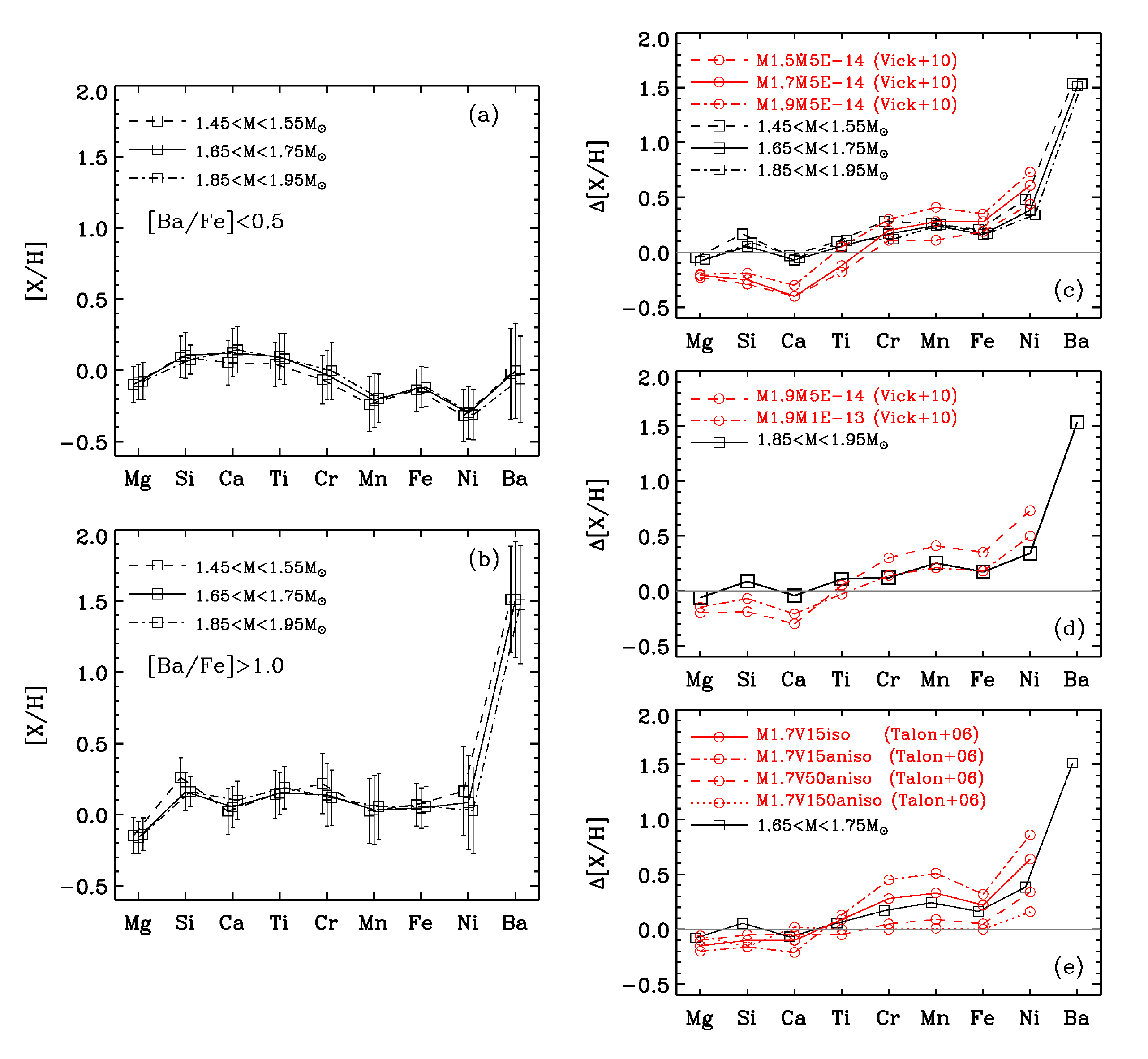}
\caption{$Left$: mean abundances and dispersion for Ba-normal stars (panel $a$) and Ba-enhanced stars (panel $b$) with different masses in the temperature range of $6700<T_{\rm eff}<7500$\,K. $Right$: Differential abundances between Ba-enhanced and Ba-normal stars (black), overplotted with the mass loss model predictions of \citet{Vick2010} for different masses (panel $c$) and different mass loss rates (panel $d$), as well as the rotation-reduced turbulence model of \citet{Talon2006} (panel $e$). In panel $c$, three models are shown corresponding to masses of 1.5, 1.7, and 1.9\,$M_\odot$ and ages of 0.5, 0.5, and 0.8\,Gyr, respectively. All models have a mass loss rate of $5\times10^{-14}$\,$M_\odot$/year. In panel $d$, the models have a fixed mass of 1.9\,$M_\odot$ but with different mass loss rates ($5\times10^{-14}$\,$M_\odot$/year versus $1\times10^{-13}$\,$M_\odot$/year). In panel $e$, all models have a fixed mass of 1.7\,$M_\odot$ and a constant age of 0.8\,Gyr, but with different rotation velocities --- $V_{\rm rot}$ = 15\,km/s, no anisotropic diffusion, $V_{\rm rot}$ = 150\,km/s, anisotropic diffusion, $V_{\rm rot}$ = 50\,km/s, anisotropic diffusion, $V_{\rm rot}$ = 15\,km/s, anisotropic diffusion; see Table\,1 and Fig.\,14 of \citet{Talon2006}.  }
\label{fig:Fig14}
\end{figure*}

Fig.\,\ref{fig:Fig14} shows the mean abundance and dispersion for both peculiar and normal stars with a range of masses and compares the differential abundances of these stars with the stellar radiative acceleration models of \citet{Vick2010} and \citet{Talon2006}. 
There we see that the mean abundances of normal stars are similar (within $\sim0.1$\,dex; e.g., panel $a$) among different masses. This suggests that the abundances generally do not vary strongly with age for these relatively young stellar populations. 
The normal stars with similar masses exhibit a dispersion of $\sim$0.2\,dex, significantly larger than the reported measurement errors ($\lesssim$0.05\,dex). This could be due to spatial variation stemming from local abundance gradients or possibly non-negligible atomic diffusion. Interestingly, the Ba-enhanced chemically peculiar stars of different masses also exhibit consistent mean abundances (e.g., panel $b$). This might be a useful constraint on the stellar physical processes causing the chemical peculiarity, such as variations in mass loss rate with stellar mass. All masses  exhibit large scatter in abundances (0.1--0.2\,dex for Mg, Si, Ca, Ti, Fe; 0.3--0.5\,dex for Cr, Mn, Ni, Ba). For a few elements, such as Cr, Mn, and Ni, the scatter is apparently larger than that of the chemically normal stars. This is probably due to the fact the abundances of these Ba-enhanced chemically peculiar stars have been altered by a variety of factors that depend, for instance, on age, mass loss rate, and rotation speed.  

The differential abundances between the peculiar and normal stars exhibit a similar pattern to predictions from the model of \citet{Vick2010}, in which surface abundances are the consequence of competition between atomic diffusion due to gravitational settling and radiative acceleration, modulated by stellar mass loss (panel $c$). However, there are significant differences between the model and the observations for a few elements. Whereas stars with M = 1.5\,$M_\odot$ and M = 1.7\,$M_\odot$ exhibit iron-peak elemental abundances (Cr, Mn, Fe, Ni) in good agreement with the models for a mass loss rate of $5\times10^{-14}M_\odot$/year, the models predict stronger depletion of Mg, Si, Ca, and Ti.  The pattern in abundances predicted for these elements, however, is similar to the observations; in both cases Si and Ti abundances are higher than Ca. For stars with M = 1.9\,$M_\odot$, the model with a mass loss rate of $1\times10^{-13
}M_\odot$/year provides a much better match to the observations than the model with a mass loss rate of $5\times10^{-14}M_\odot$/year (panel $d$). 

The observed differential abundance patterns are also qualitatively consistent with the model of \citet{Talon2006}, in which the surface abundances are modulated by turbulence induced by stellar rotation (panel $e$). The values of the abundance alteration depend sensitively on the rotation velocity $V_{\rm rot}$. The observed differential abundances are comparable to the prediction of an isotropic diffusion model with $V_{\rm rot}$ of about 15\,km/s or an anisotropic diffusion model with $V_{\rm rot}$ between 15\,km/s and 50\,km/s. These rotation velocities are qualitatively consistent with the observed $v{\rm sin}i$ from GALAH DR2 (Fig.\,\ref{fig:Fig12}). For Mg, Si, Ca, and Ti, the turbulence models seem to match the data better than the mass loss models. This is consistent with the finding of \citet{Michaud2011}. 

These results suggest the Ba-enhanced chemically peculiar stars are inevitably the consequence of stellar internal atomic transport processes. We cannot, however, strongly prefer one model over the other. Indeed, there are a number of reasons why we might expect the models and the observations to differ. First, as mentioned above, the ``normal" stars may not actually reflect true initial abundances. Second, the mass loss rate adopted by the model may not be the optimal one, and it is possible that stars with the same mass exhibit different mass loss rates, depending on a number of factors, such as age and rotation speed. Furthermore, it is possible that both processes, turbulence and mass loss, are present in stars.  Understanding how these two processes are related and act together is key from both observational and theoretical points of view.  

\subsection{External accretion}
Stellar photospheric abundances can also be altered by external accretion events, such as supernova (SN) pollution, mass transfer from an AGB companion, and planet engulfment. In this section, we briefly discuss the potential impact (if any) of these processes on the origin of these A and F-type Ba-enhanced chemically peculiar stars.
\subsubsection{Supernova pollution}
The origin of the Ba-enhanced chemically peculiar stars are hard to explain by pollution from supernova companions, both due to their high incidence and their characteristic abundance patterns.  Core-collapse supernovae produce a high fraction of $\alpha$-elements, for example, making it unlikely that they would yield the observed low abundances of Mg and Ca.  The high incidence of the Ba-enhanced chemically peculiar stars also seems to rule out explosive (Ia) supernovae.   As shown in Section\,3.3, about 40\% stars with mass larger than 1.5\,$M_\odot$ are found to exhibit peculiar abundances, and the fraction could even be higher if we consider that there may be even more peculiar stars with ${\rm [Ba/Fe]}<1$ that are not accounted for. However, only 1.5--2 percent of white dwarfs (WDs) will become SN Ia \citep{Maoz2018}, which means that even if we assume that all stars with $6700<T_{\rm eff}<7500$ have WD companions, we should only see 2\% stars with peculiar abundance patterns. In addition, in order to have a SN Ia companion, an A-type star would likely need to be born in a triple system, wherein the two more massive siblings formed the SN Ia companion. This further reduces the frequency of such an event by a factor of 5 \citep{Duchene2013}. Supernova Ia are also not expected to produce stars with the high [Ba/Fe] values observed. 

\subsubsection{Mass transfer from AGB companion}
The over-abundances could be due to the accretion of material from AGB companions, which produce a large amount of Ba \citep[e.g.,][]{Busso1999, Karakas2014}.  In this case, we might expect the enhanced stars to hold unseen white dwarf (WD) companions that have evolved from progenitor AGB stars.  To further consider this possibility, here we examine the observed kinematics and ultraviolet (UV) properties of the Ba-enhanced stars for signals of WD companions.

First, we examine the scatter in radial velocities (RV) between multi-epoch observations from both \textit{Gaia} DR2 and LAMOST for signals of binarity. 
For \textit{Gaia} DR2, the RV scatter is derived by multiplying the reported radial velocity errors by the number of measurements, given that the former is defined as the standard deviation of multi-epoch radial velocity measurements \citep[see Eq.\,1 of ][]{Katz2019}. For LAMOST, about one third of the whole LAMOST spectral dataset are repeat visits of common targets, and for each visit, there are two or three consecutive exposures \citep[for the exposure strategy of LAMOST, see, e.g.,][]{Yuan2015}. Recently, radial velocities have been derived from these single-exposure spectra (Yang et al. 2020, in preparation) using the LAMOST stellar parameter pipeline at Peking University \citep[LSP3;][]{Xiang2015}. Uncertainties in the radial velocity measurements depend on both the apparent magnitude and the spectral type of each star. To ensure sufficient radial velocity precision, we select only Gaia DR2 sample stars with G-band magnitude brighter than 12\,mag, and LAMOST sample stars with $S/N>50$. For these relatively hot A/F-type stars, the typical radial velocity error for a $single$-epoch measurement is expected to be 4--5\,km/s for both Gaia DR2 and LAMOST. Note that this is $\sim$\,$\sqrt{\pi/2N}$ times larger than the reported Gaia radial velocity uncertainty, which is for the combined (mean) velocity, where $N$ represents the number of observation epochs. We further require that each target is observed in at least five epochs, leading to a total of 4002 stars from \textit{Gaia} DR2 and 5276 stars from LAMOST with $6700<T_{\rm eff}<7500$\,K and ${\rm [Fe/H]}>-0.2$\,dex.  The RV variations measured with LAMOST are quantified as the ratio between the scatter of the radial velocity measurements and the measurement error, $R = \sqrt{\Sigma_{i=1}^N(v_i-\bar{v})^2 / \sigma_{v,i}^2}/\sqrt{N}$, where $v_i$ is the velocity of the $i_{\rm th}$ epoch and $\sigma_{v,i}$ is the velocity error.

For a binary system in a Keplerian orbit, the velocity amplitude of the A/F star's companion would be \[K_1 = \sqrt{GM_2^2/[a(M_1+M_2)(1-e^2)]} \times{\rm sin}i, \] where $M_1$ and $M_2$ are the masses of the observed A/F star and the unseen WD, respectively, $a$ is the total separation, $e$ is the orbital eccentricity, and $i$ is the inclination angle of the system.
Here we consider $M_1=1.5$\,M$_\odot$ and adopt $M_2$ = 0.6\,M$_\odot$ typical of DA white dwarfs \citep[e.g.,][]{Rebassa-Mansergas2015}.  We also assume that $e = 0$, which is plausible for a close binary, leading to \[K_1=\frac{12.3}{\sqrt{a({\rm AU})}}{\rm sin}i ~ ({\rm km\,s}^{-1}).\] Given the precision of radial velocities in \textit{Gaia} DR2 and LAMOST discussed above, this implies that unseen WD companions should be identifiable within about 10\,AU of the target stars.  About half of the discovered binary systems match this criterion \citep[see Fig.\,2 of][]{Duchene2013}. 
     
\begin{figure*}[ht!]
\centering
\includegraphics[width=180mm]{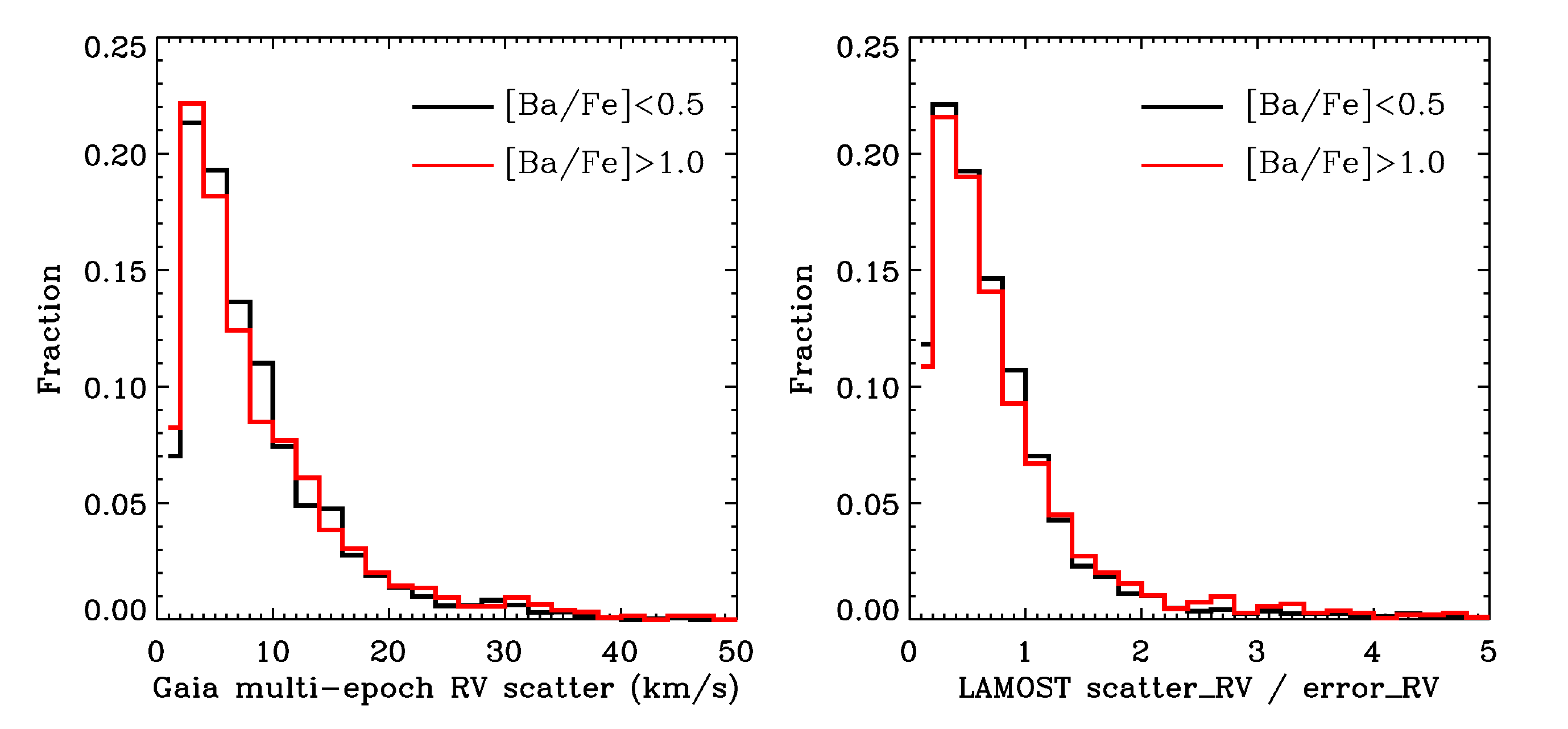}
\caption{$Left:$ Scatter in \textit{Gaia} multi-epoch radial velocity measurements for Ba-normal (black) and Ba-enhanced (red) stars with $6700<T_{\rm eff}<7500$\,K and ${\rm [Fe/H]}>-0.2$\,dex. The \textit{Gaia} G mag is restricted to be brighter than 12\,mag to ensure precise radial velocity measurements. $Right:$ Radial velocity variations from LAMOST multi-epoch observations. The variation is quantified by the ratio of the scatter in radial velocity measurements and the measurement error, i.e. $R = \sqrt{\Sigma_{i=1}^N(v_i-\bar{v})^2 / \sigma_{v,i}^2}/\sqrt{N}$, where $v_i$ is the velocity of the $i_{\rm th}$ epoch, $\sigma_{v,i}$ is the velocity error, and $N$ is the total number of observation epochs.  The stars in both samples are required to have been observed in at least five epochs.   The distributions of Ba-normal and Ba-enhanced stars exhibit no obvious differences.}
\label{fig:Fig15}
\end{figure*}

Fig.\,\ref{fig:Fig15} compares the distribution in RV scatter between multi-epoch measurements for the Ba-enhanced chemically peculiar stars and for  chemically normal stars, measured with both \textit{Gaia} DR2 and LAMOST. Chemically peculiar stars do not exhibit significantly greater RV scatter than chemically normal stars, implying that binary evolution is unlikely an important driver of chemical peculiarity. This finding is consistent with previous results for Ba-rich main sequence (turnoff) stars \citep{Milliman2015}, but is notably different than appears to be the case for Ba-rich giant stars, which are widely believed to be binary products \citep{McClure1983}. 
 
\begin{figure}[ht!]
\centering
\includegraphics[width=85mm]{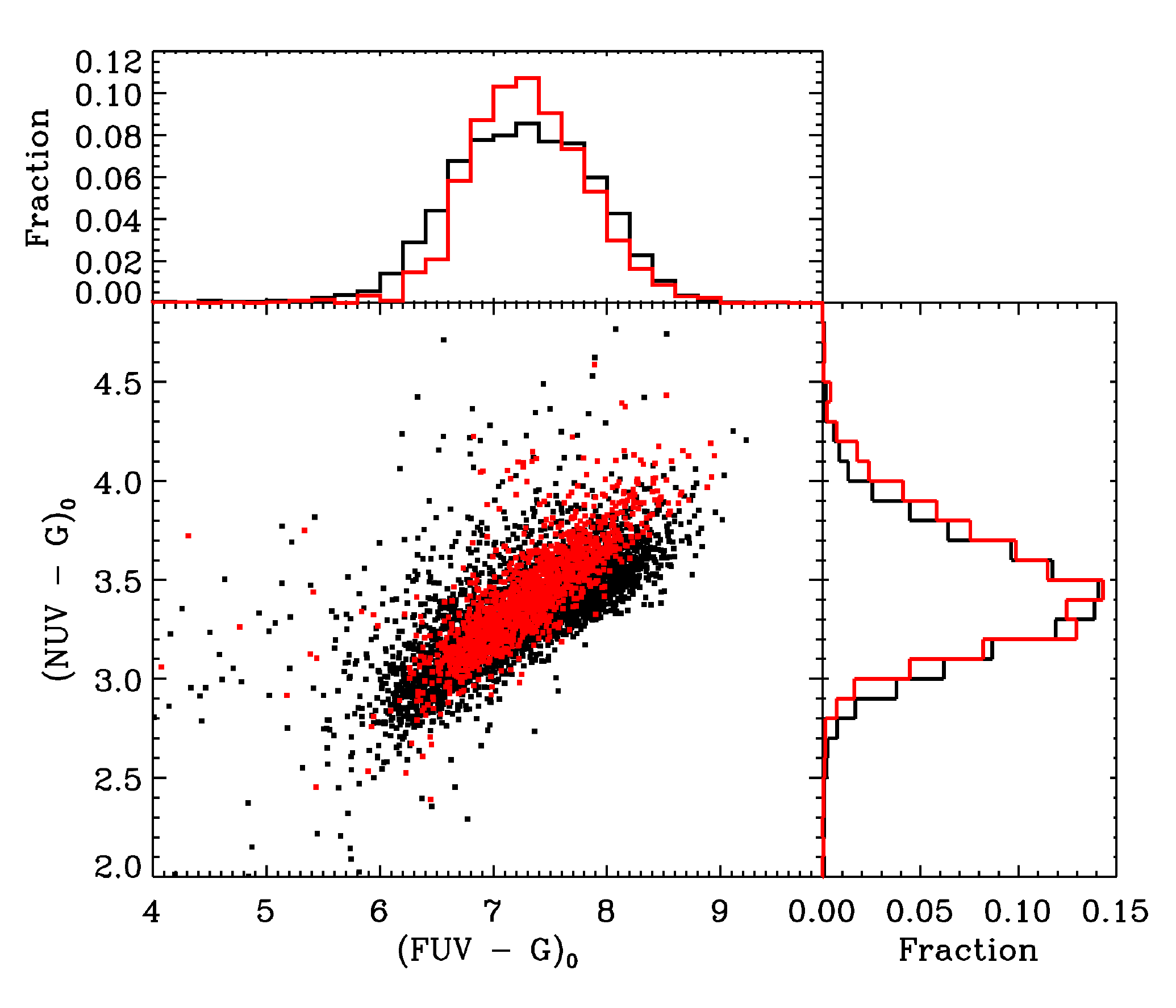}
\caption{Comparison of chemically peculiar (red) and normal (black) stars in the $FUV-G$ versus $NUV-G$ color diagram. The $FUV$ and $NUV$ photometry are from the Galex DR5, and the $G$-band magnitude is from Gaia DR2. A photometric error of better than 0.1\,mag for both FUV and NUV magnitudes is required.}
\label{fig:Fig16}
\end{figure}

WDs can also reveal themselves through differences in the $UV$ -- optical colors of chemically peculiar stars vs. chemically normal stars.  To construct  color measurements we use the far-$UV$ ($FUV$) and near-$UV$ ($NUV$) photometry from the Galex DR5 \citep{Bianchi2011}, for which a photometric error of better than 0.1\,mag is required for both filters, and use the $G$ magnitude from the \textit{Gaia} DR2 \citep{Brown2018}.  
The photometry is dereddened using $E(B-V)$ derived from the star pair method \citep{Yuan2013}, which takes as input the LAMOST {\it DD-Payne} stellar parameters and multi-band photometry (a brief introduction can be found in \citealt{Xiang2019}). The typical precision of the $E(B-V)$ estimates is 0.01--0.02\,mag. To convert $E(B-V)$ to reddening in the Galex and \textit{Gaia} passbands, we use an extinction coefficient depending on $T_{\rm eff}$ and $E(B-V)$ that we have computed by convolving the Kurucz synthetic spectra \citep{Castelli2003} with the Fitzpatrick extinction law \citep{Fitzpatrick1999}. Fig.\,\ref{fig:Fig16} shows the distributions of chemically peculiar and chemically normal stars in the $FUV-G$ versus $NUV-G$ color-color diagram. 

For reference, we have examined Gaia WDs within 300\,pc of the Solar Neighbourhood, finding an average $FUV$ absolute magnitude of 11.0\,mag, which is only 0.6\,mag drimmer than the $FUV$ absolute magnitude of our A/F sample stars ($6700<T_{\rm eff}<7500$\,K) within 300\,pc. In the Gaia G-band, these WDs are much fainter (by 8.5\,mag) than A/F stars. This suggests that, if there is a white dwarf (WD) companion around a Ba-enhanced chemically peculiar star, we should observe an $FUV-G$ color excess of 0.5\,mag. However, Fig.\,\ref{fig:Fig16} demonstrates that the chemically peculiar stars have similar colors to normal stars, implying that they likely do not have WD companions.

The significant enhancement in iron-peak elements but depletion in Mg characteristic of Ba-enhanced chemically peculiar stars also strongly argues against mass transfer from AGB companions as the responsible mechanism, given that AGB stars do not produce such an abundance pattern.

\subsubsection{Planet engulfment}
It has been suggested that earth-like planets could be engulfed onto the surface of a main sequence (turnoff) star, altering the observed chemical stellar surface abundances as the engulfed planet gets dissolved and mixed in the convective envelope over short time scales (a few years) \citep{Church2019}. To investigate if this mechanism can provide a reasonable explanation for Ba-enhanced chemically peculiar stars, we carried out rough estimates of the abundance pattern of an A/F star assuming it has engulfed a terrestrial planet with chemical composition similar to either the Earth or Mercury into its convective envelope (since the chemically peculiar stars are iron-peak enhanced, if they formed via planet engulfment, the engulfed planets would seem most plausibly terrestrial ones). For element $X$, the minimum mass that must have been accreted from the planet in order to enrich the stellar envelope from an initial abundance [$X$/H]$_0$ to a current abundance [$X$/H] is
\begin{equation}
M_{X,{\rm planet}} = M_{\rm cenv} \times X_\odot/H_\odot \times \left(10^{[X/H]}-10^{[X/H]_0}\right) \times A_X
\end{equation}
where $M_{\rm cenv}$ is the mass of the stellar convective envelope, $X_\odot/H_\odot$ is the solar abundance of $X$ in absolute value, and $A_X$ is the atomic mass number of $X$. This assumes that all accreted material is reserved in the convective envelope, although part of this material in reality is likely to enter below the base of the convective envelope, in which case planets with larger masses than considered here would be required. Our fiducial case adopts a star with a convective envelope mass of $1.5\times10^{-4}$\,$M_{\odot}$, typical for stars at the border of the $T_{\rm eff}$--$\log g$ diagram (Fig.\,\ref{fig:Fig7}), and an initial abundance (before accreting the planet) set to the median values of the abundances characteristic of chemically normal stars. We adopt the estimates of \citet{Morgan1980} for the abundances of Mercury and adopt abundances from \citet{McDonough1995} for the Earth. 

\begin{figure}[htp]
\centering
\includegraphics[width=85mm]{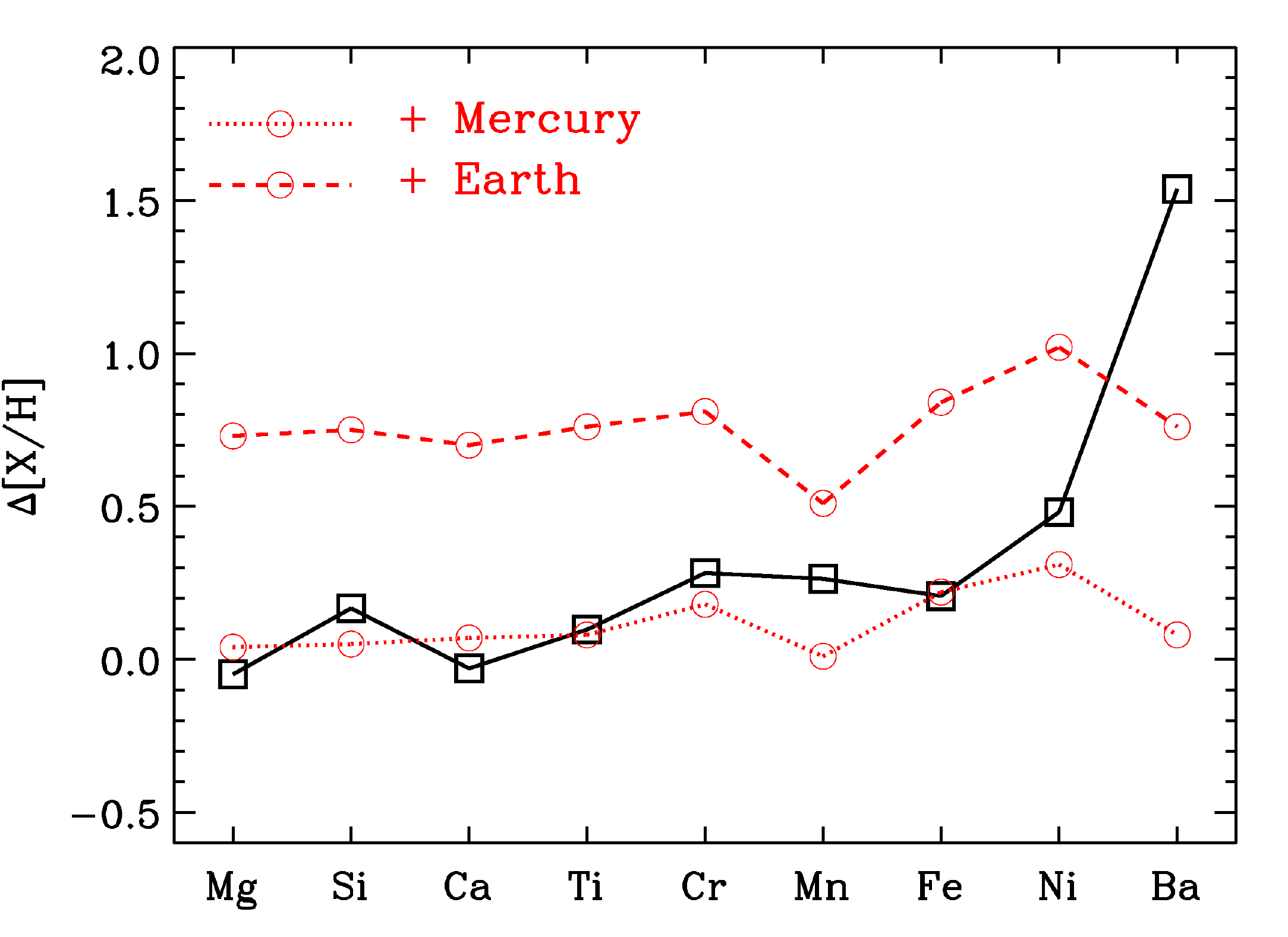}
\caption{Abundance alteration for a 1.5\,$M_\odot$ star after accreting a Mercury-like planet (red dotted line) and an Earth-like planet (red dashed line). The black line shows the observed differential abundances between 
chemically peculiar and normal stars with $M=1.5$\,$M_\odot$. }
\label{fig:Fig17}
\end{figure}

Fig.\,\ref{fig:Fig17} shows the alteration in stellar abundances for the fiducial star after accreting a Mercury or an Earth.   The observed abundance enhancement/depletion patterns of the Ba-enhanced chemically peculiar stars are shown for comparison, adopting the differential abundances between chemically peculiar and normal stars, as in Section\,\ref{radiativeacceleration}.  Ba-enhanced chemically peculiar stars exhibit much stronger Ba enhancement than caused by the engulfment of a Mercury or an Earth-like planet. The depletion of Mg and Ca characteristic of Ba-enhanced chemically peculiar stars is also inconsistent with the engulfment of terrestrial planets like the Earth and Mercury, given that they contain large amounts of these elements and will cause a significant abundance enhancement to the star after the engulfment events.  At the same time, these planets can not be responsible for the enhancements in Cr, Mn, and Ni observed for the Ba-enhanced stars. These patterns are more consistent with enhancement via a stellar evolution mechanism, as discussed in the last section. We therefore conclude that planet engulfment is quite an unlikely explanation, with the caveat that the details of how mixing takes place in the convective envelope may be important.  It has been suggested that the material accreted onto the surface of a star could be significantly reduced in a short time scale (1000 years) due to thermohaline mixing \citep{Vauclair2004, Theado2009, Theado2012}. 

\section{Conclusion} \label{conclusion}
We have identified and analyzed 15,009 metal-rich (${\rm [Fe/H]}>-0.2$\,dex) and Ba-enhanced (${\rm [Ba/Fe]}>1$\,dex) stars from the LAMOST DR5 abundance catalog of \citet{Xiang2019}. We find that they are dominated by relatively hot main-sequence and subgiant stars with $T_{\rm eff}\gtrsim6300$\,K and that their distribution in the $T_{\rm eff}$ -- $\log g$ diagram exhibits a sharp border at the lower-$T_{\rm eff}$ side, corresponding to an approximately fixed mass of 1.4\,$M_\odot$ or a fixed convective envelope mass $10^{-4}$ times that of the star. Statistically, these Ba-enhanced stars exhibit enhanced abundances for all the iron-peak elements (Cr, Mn, Fe, Ni) compared to the Ba-normal (${\rm [Ba/Fe]}<0.5$\,dex) stars, but depleted abundances of Mg and Ca. These characteristics suggest they are likely related to the Am/Fm stars that have been found since 1930's. Comparisons of these abundance patterns with stellar evolution models that account for radiative acceleration and stellar rotation (or mass loss) show good consistency, suggesting these Ba-enhanced chemically peculiar stars are consequences of stellar evolution, i.e., the competition between radiative acceleration and gravitational settling. Ba-enhanced chemically peculiar stars generally exhibit lower rotation velocities $v{\rm sin}i$ (taken from GALAH DR2) than chemically normal stars. These metal-rich, Ba-enhanced chemically peculiar stars constitute about 16\% of the whole stellar population in the temperature range $6700<T_{\rm eff}<7500$\,K, and this fraction reaches as high as 40\% for stars more massive than 1.5\,$M_\odot$, suggesting that ``peculiar" photospheric abundances are a ubiquitous phenomenon for these intermediate-mass stars. These results call for cautious treatment when employing intermediate-mass stars for a variety of purposes, e.g., studying Galactic chemical evolution, deriving stellar ages with isochrones, etc.

\vspace{7mm} \noindent {\bf Acknowledgments}{
The authors thank Andy Gould, Jeffrey Gerber, Karin Lind and Xian-Fei Zhang for helpful discussions. We are grateful for the anonymous referee for insightful comments and suggestions, and Dr. Sharon van der Wel for her careful proofreading and editing on the manuscript. H.-W. Rix and H.-G. Ludwig acknowledge funding by the Deutsche Forschungsgemeinschaft (DFG, German Research Foundation) -- Project-ID 138713538 -- SFB 881 (``The Milky Way System'', subproject A03 \& A04). Y.S.T. is grateful to be supported by the NASA Hubble Fellowship grant HST-HF2-51425.001 awarded by the Space Telescope Science Institute. H.-W. Zhang and M. Zhang acknowledge the National Natural Science Foundation of China 11973001 and National Key R\&D Program of China No. 2019YFA0405504. S. Buder acknowledges funds from the Alexander von Humboldt Foundation in the framework of the Sofja Kovalevskaja Award endowed by the Federal Ministry of Education and Research as well as support by the Australian Research Council (grants DP150100250 and DP160103747). Parts of this research were supported by the Australian Research Council (ARC) Centre of Excellence for All Sky Astrophysics in 3 Dimensions (ASTRO 3D), through project number CE170100013.}

\bibliographystyle{mn2e}
\bibliography{reference.bib}

\end{document}